\documentclass[
  journal=pasa,
  manuscript=research-article,
  year=2026,
  volume=37,
]{cup-journal}

\usepackage{amsmath}
\usepackage[nopatch]{microtype}
\usepackage{booktabs}
\usepackage{subfigure}
\usepackage{placeins}
\usepackage{orcidlink}
\hypersetup{hidelinks}
\usepackage{amssymb}

\title{The TASSIE Program. II: Three Close-In Companions Orbiting Sun-Like Stars}

\author{T. Plunkett \orcidlink{0009-0003-5810-1314}}
\affiliation{Greenhill Observatory, School of Natural Sciences, University of Tasmania, Private Bag 37, Hobart, TAS 7001 Australia}
\email[T. Plunkett]{thomas.plunkett@utas.edu.au }

\author{E. Thygesen \orcidlink{0000-0002-9165-6245}}
\affiliation{Greenhill Observatory, School of Natural Sciences, University of Tasmania, Private Bag 37, Hobart, TAS 7001 Australia}

\author{A. A. Cole \orcidlink{0000-0003-0303-3855}}
\affiliation{Greenhill Observatory, School of Natural Sciences, University of Tasmania, Private Bag 37, Hobart, TAS 7001 Australia}

\author{J. Schulte \orcidlink{0000-0002-7382-0160}}
\affiliation{Center for Data Intensive and Time Domain Astronomy, Department of Physics and Astronomy, Michigan State University, East
Lansing, MI 48824, USA}

\author{J. E. Rodriguez \orcidlink{0000-0001-8812-0565}}
\affiliation{Center for Data Intensive and Time Domain Astronomy, Department of Physics and Astronomy, Michigan State University, East
Lansing, MI 48824, USA}

\author{B. Emptage}
\affiliation{Greenhill Observatory, School of Natural Sciences, University of Tasmania, Private Bag 37, Hobart, TAS 7001 Australia}

\author{J. P. Beaulieu \orcidlink{0000-0003-0014-3354}}
\affiliation{Greenhill Observatory, School of Natural Sciences, University of Tasmania, Private Bag 37, Hobart, TAS 7001 Australia}
\alsoaffiliation{Sorbonne Universit\'e, CNRS, Institut d'Astrophysique de Paris, IAP, F-75014, Paris, France}

\author{G. Bakos \orcidlink{0000-0001-7204-6727}}
\affiliation{Department of Astrophysical Sciences, Princeton University, Princeton, NJ 08544, USA}

\author{J. Hartman \orcidlink{0000-0001-8732-6166}}
\affiliation{Department of Astrophysical Sciences, Princeton University, Princeton, NJ 08544, USA}

\author{C. Ziegler \orcidlink{0000-0002-0619-7639}}
\affiliation{Department of Physics, Engineering and Astronomy, Stephen F. Austin State University, 1936 North St, Nacogdoches, TX 75962, USA}

\author{K. A. Collins}
\affiliation{Center for Astrophysics, Harvard \& Smithsonian, 60 Garden Street, MS 78
Cambridge, MA 02138, USA}

\author{Z. Csubry \orcidlink{0000-0002-8423-0510}}
\affiliation{Department of Astrophysical Sciences, Princeton University, Princeton, NJ 08544, USA}

\author{K. Penev \orcidlink{0000-0003-4464-1371}}
\affiliation{Department of Physics, University of Texas at Dallas, 800 W. Campbell Road, Richardson, TX 75080, USA}

\author{A. Jord\'an \orcidlink{0000-0002-5389-3944}}
\affiliation{Facultad de Ingenier\'ia y Ciencias, Universidad Adolfo Ib\'{a}\~{n}ez, Av. Diagonal las Torres 2640, 7941169 Pe\~{n}alol\'{e}n, Santiago, Chile}
\alsoaffiliation{Departamento de Astronomía, Universidad de Chile, Casilla 36-D, Santiago, Chile}

\author{R. Brahm}
\affiliation{Facultad de Ingenier\'ia y Ciencias, Universidad Adolfo Ib\'{a}\~{n}ez, Av. Diagonal las Torres 2640, 7941169 Pe\~{n}alol\'{e}n, Santiago, Chile}

\author{L. Mancini \orcidlink{0000-0002-9428-8732}}
\affiliation{Department of Physics, University of Rome ``Tor Vergata'', Via
della Ricerca Scientifica 1, 00133 Rome, Italy}
\alsoaffiliation{INAF -- Turin Astrophysical Observatory, via Osservatorio 20,
10025 Pino Torinese, Italy}
\alsoaffiliation{Max Planck Institute for Astronomy, K\"{o}nigstuhl 17, 69117
Heidelberg, Germany}

\author{T. Henning}
\affiliation{Max Planck Institute for Astronomy, K\"{o}nigstuhl 17, 69117
Heidelberg, Germany}

\author{D. J. Radford \orcidlink{0000-0002-3940-2360}}
\affiliation{Brierfield Observatory, Bowral, NSW Australia}

\author{P. Evans \orcidlink{0000-0002-5674-2404}}
\affiliation{El Sauce Observatory, Coquimbo Province, Chile}

\keywords{exoplanets, transits, telescopes} 

\begin{document}
\newcommand{\gaia}{\textit{Gaia\ }}

\begin{abstract}
We present three southern transiting giant planet candidates alerted by the Transiting Exoplanet Survey Satellite (\textit{TESS}) mission and investigated at the University of Tasmania Greenhill Observatory (UTGO). The candidate planets are orbiting thin disk G-dwarf main-sequence stars with roughly solar metallicity, possessing orbital periods between 2.9 - 3.3 days and radii of 1.1 - 1.3 $R_{J}$. We performed ground-based follow-up photometry primarily with the UTGO Harlingten 50 cm, then gathered reconnaissance spectra, high angular resolution imaging and high-precision radial velocities to rule out false positive scenarios. We confirmed that two of these systems host true exoplanets and constrained their masses. TOI-3053b is a typical hot Jupiter, with $M_{3053b} = 0.85 \pm 0.12$ M$_{J}$ and a bulk density of $\rho_{3053b} = 0.64 \pm 0.10$ g~cm$^{-3}$. TOI-3278b / HATS-78b is a hot Saturn-mass planet ($M_{3278b} = 0.30 \pm 0.07$ $M_{J}$) with a highly inflated atmosphere and a low density of $\rho_{3278b} = 0.21 \pm 0.05$ g~cm$^{-3}$. The other candidate (TOI-3272.01) remains unconfirmed, but appears consistent with being a hot Jupiter. TOI-3272.01 is notable as a candidate planet orbiting a potentially young to intermediate age star, with a rotational analysis indicating an age estimate of  $T_{3272} = 1.1 \pm 0.2$ Gyr. These systems add to a growing sample of hot giant planets from \textit{TESS} that may provide constraints on the migration pathways and radius inflation of the broader close-in exoplanet population.   
\end{abstract}

\section{Introduction}
Despite being the first type of exoplanet to be discovered around a main-sequence star \citep{mayorqueloz}, hot giant planets (M$_{p}$ $\gtrsim$ 0.3 $M_{J}$, P < 10 days) continue to challenge our understanding of planet formation and evolution. Although it is widely accepted that orbital migration is necessary to move these planets from their formation locations beyond the snowline to their current close-in orbits, the exact mechanisms are still contested \citep{origins}. The two predominant theories are high-eccentricity migration (HEM) and disk-driven migration. In the first scenario, planets are perturbed to an eccentric orbit by the gravitational influence of an outer body (such as another planet or a stellar companion, known as Kozai-Lidov oscillations), secular interactions or planet-planet scattering \citep[e.g.,][]{Naoz2016, Adams2006, Ford2008}. Tidal interactions with the host star during close approaches at periastron then circularise and shrink the orbit. Whilst the majority of hot giants are observed to have circular orbits, a non-negligible fraction remain eccentric (i.e., are in the midst of this process) with a distribution consistent with HEM \citep[e.g.,][]{bonomo2017}. For instance, many systems hosting hot Jupiters (HJs) have also been found to contain outer gas giant companions, with their HJs possessing higher average eccentricities than those in single planet systems \citep{Knutson2014, Bryan2016}. Furthermore, the shape of the upper boundary of the sub-Jovian desert (a dearth of giants in a region of the close-in planet parameter space) also lends support to the HEM theory \citep[see][]{owenlai2018}.
\newline\newline 
The second scenario of disk-driven migration results from the exchange of angular momentum and energy between an embedded planet and protoplanetary disk \citep{goldreich1980}. Depending upon the presence of gaps in the disk, along with the viscosity and scale height, this mechanism can result in the rapid inward (or outward) migration of a planet \citep[see][]{diskmigration}. Disk-driven migration may be more relevant to circularised and mildly eccentric warm Jupiters (WJs, 10 < P < 200 days), as it is not fully consistent with the mass and eccentricity distributions of HJs observed \citep{warmsecular}. However, the existence of multi-planet systems with close companions to HJs indicates that this process must occur in some cases, as a highly dynamical pathway such as HEM would disrupt the architecture of these systems \citep[e.g.,][]{Becker2015, Huang2020}. It may be that both migration mechanisms are required (with differing probabilities) to reproduce the trends observed in the short-period giant population \citep{origins}.
\newline\newline
Since the launch of the Transiting Exoplanet Survey Satellite \citep[\textit{TESS},][]{tess} in 2018, hundreds of new close-in giant planet candidates have been announced around the Solar neighbourhood (d < 1 kpc). With an increasing number of well-characterised giants with accurate host star parameters, we can now assess the influence of host stars on the evolution of this class of planets. Of particular interest are systems with young to intermediate age stars ($T < 1$ Gyr) and post-main sequence stars, as these populations were sparsely studied in the era of the \textit{Kepler} mission \citep{kepler} and are vital for understanding the radius inflation of HJs and other close-in giants \citep[e.g.,][]{weiss2013, inflation}. Comparisons between the occurrence rates of giant planets around different stellar populations (such as low-mass, thin/thick disk, halo and those in stellar clusters) may also expose the circumstances of their formation and evolution in greater detail \citep[e.g.,][]{thyme, gems}. Furthermore, the James Webb Space Telescope \citep[\textit{JWST},][]{JWST} and the upcoming \textit{Ariel} mission \citep{ariel} will allow us to probe the atmospheric compositions of a large sample of giant planets through transmission spectroscopy. Certain chemical signatures, such as the C/O ratio (a tracer of birth location), may remain from formation and allow us to discriminate between possible origin scenarios \citep[see][]{mordasini2016, hotjupariel}. Hence, it remains vital to perform follow-up on giant planet candidates from \textit{TESS} in order to validate and characterise interesting targets for these missions. 
\newline\newline 
The `TASmanian Search for Inclined Exoplanets' (TASSIE) program aims to study transiting (i.e., inclined) planets in the sub-Jovian desert alerted by the \textit{TESS} mission, along with other hot giants in the southern sky \citep[for further details, see][]{tassie}. In this work, we present three new giant planet candidates alerted by \textit{TESS} and investigated with the Harlingten 50 cm telescope at University of Tasmania Greenhill Observatory (UTGO). The paper is organised as follows. In Section 2, we detail the observations of these system including ground-based follow-up photometry from UTGO, reconnaissance spectroscopy, high angular resolution speckle imaging and high-precision RV measurements. In Section 3, we derive the stellar characteristics from spectroscopic analysis and spectral energy distribution fitting. We then move on to planet validation in Section 4 using archival imaging, astrometry and statistical testing with TRICERATOPS \citep{triceratops}. We then perform joint modelling of the light curves and RVs with Juliet \citep{juliet}. In Section 5, we discuss the three systems in the context of the short-period planet population and suggest future observations. Finally, we present our conclusions in Section 6. 

\section{Observations}
As part of the TASSIE program, these targets were selected for follow-up from the \textit{TESS} Object of Interest (TOI) sample following the criteria outlined in \citet{tassie}. We now detail the observations collected for each system. 

\subsection{Photometry}
We obtained the \textit{TESS} data, including the Target Pixel Files (TPFs), Full Frame Images (FFIs) and light curves from the Mikulski Archive for Space Telescopes (MAST). For two of the targets, we found that the light curves for some sectors were unavailable (or were from inconsistent pipelines) when searching through the Lightkurve package \citep{lightkurve}. To ensure that the light curve analysis and derived transit depths were unbiased by these differing extractions, we decided to create our own light curves for two of the targets from their FFIs using the TESS-GAIA Light Curve (TGLC) package \citep{tglc}. This package utilises \gaia DR3 \citep{dr3} photometry and astrometry to forward model the FFIs, building a per-image effective point spread function (ePSF). This has been shown to produce better photometric precision and more accurate de-blending, especially for faint targets.
\newline\newline
We performed ground-based follow-up observations primarily with the Harlingten 50 cm (H50) telescope at the University of Tasmania Greenhill Observatory (UTGO). This was done to confirm the source of the transit, rule out false positives (such as eclipsing binaries and stellar activity) and refine the orbital ephemerides. The H50 is equipped with an Apogee ALTA U42 CCD camera, with a pixel scale of 0.8 arcsec~pix$^{-1}$ and a 27.2$^{\prime}$$\times$27.2$^{\prime}$ field-of-view. Data were reduced and differential photometry was extracted by a custom pipeline utilising the Prose package \citep{prose}. The logs of the H50 observations are shown in Table \ref{table:tassielog}, with logs for the additional ground-based photometry in the Appendix (Table \ref{table:tassiephotlog2}). We now describe the observations for each target in greater detail.

\subsubsection{TIC 269859655 - TOI-3053}
TIC 269859655 is a V $\approx$ 13.2 mag star in the constellation of Carina. It was observed by \textit{TESS} in Sectors 9, 10, 11, 36, 37, 38, 64 and 90 at cadences ranging from 1800 seconds down to 200 seconds. Raw photometric data were processed by the TESS Quick Look Pipeline \citep[QLP,][]{QLP}, which performs difference imaging photometry, preliminary de-trending of instrumental effects and flux de-blending. TOI-3053 was originally flagged as a community TOI in \citet{CTOI} and then recovered by the TESS Faint Star Search \citep{faintsearch}, with a candidate signal found at a period of 2.991 days and a large depth of 17.1 parts per thousand (ppt). We validate this signal with our own Box Least Squares \citep[BLS,][]{BLS} search, but also found an additional candidate signal at $\approx$ 0.416 days with a depth of $\approx$ 2 ppt. However, after performing a difference image centroid test\footnote{Using this package: \url{https://github.com/stevepur/transit-diffImage}}, we determined the signal was offset from the target star by roughly 17$^{\prime\prime}$ (as shown in Figure \ref{fig:centroid}). This likely indicates that this short-period signal is a blended eclipsing binary that is aligned to this target. 
\newline\newline
We observed one nearly complete transit of TOI-3053.01 on 2024/12/15 with exposures of 120 seconds in the SDSS $r^{\prime}$ filter. Imaging was disrupted by a potential time critical transient, but this alert was soon retracted and observations recommenced. The conditions were clear with moderate winds at 13 km~h$^{-1}$. We observed TOI-3053.01 again on 2025/11/19, 2025/11/22 and 2025/12/03 with the SDSS $g^{\prime}$, Bessell V and SDSS $i^{\prime}$ filters, respectively, to check the transit depth was achromatic (discussed further in Section 4.1.2). Further information on each session is shown in Table \ref{table:tassielog}. To confirm that the 0.416 day signal was a false positive, we also observed during the potential eclipse times on 2025/02/22 (assuming the full period) and 2025/03/04 (assuming half the period). We found a deep 20 ppt dip on a close F-dwarf in the first epoch, consistent with a de-blended depth estimate from the \textit{TESS} light curves and \gaia DR3 fluxes. We therefore rule this out as a signal and treat this as a single candidate system. 
\newline\newline 
Two additional light curves were obtained for TOI-3053, as part of the TESS Follow-up Observing Program Sub Group 1\footnote{\url{https://tess.mit.edu/followup/}} \citep[TFOP SG1,][]{collins2019}. One full transit was observed on 2021/09/08 using the 0.36 m telescope at Brierfield Observatory near Bowral, NSW in Australia. Images were acquired in a Johnson-Cousins R filter with exposures of 240 s using a 4096 × 4096 Moravian 16803 camera (2 x 2 binning, 1.47 arsec~pix$^{-1}$). Another full transit was observed on 2022/02/14 using the Evans 0.51m telescope at El Sauce Observatory in Coquimbo Province, Chile. 93 images were obtained with a 1536 x 1024 STT 1603-3 CCD camera (2x2 binning, 1.08 arsec~pix$^{-1}$) using exposures of 180 s in the Johnson-Cousins B filter. Photometric data from these two sessions were extracted using AstroImageJ \citep{astroimagej}.

\subsubsection{TIC 388280249 - TOI-3272}
TIC 388280249 is a V $\approx$ 14.2 mag star in the constellation of Octans. It was observed by \textit{TESS} in Sectors 11, 12, 13, 27, 38, 39, 65, 66, 67, 93 and 94 at cadences from 1800 seconds to 200 seconds. The candidate transit signal (TOI-3272.01) was identified by the Faint Star Search, with an estimated period of 3.147 days and depth of 7.4 ppt. For this target, we use the light curves extracted with the TGLC package. We identify strong photometric variability in all sectors, which is discussed in further detail in Section 3.4. 
\newline
TOI-3272 was observed by the H50 telescope on three occasions. On the first night (2024/10/02), observations were taken in the SDSS $r^{\prime}$ filter with exposures of 120 seconds. We aquired 75 images, covering the ingress, egress, and post-transit baseline. The conditions were clear with low wind. For the second night (2025/03/24), 84 images were obtained in the Bessell V filter with exposures of 120 seconds. These data covered the same transit phases as the previous observation of the target. The conditions were clear with low winds again, but the sky background was increased due to an aurora during the observations. Finally, a partial transit in the SDSS $i^{\prime}$ filter was captured on 2025/08/19, covering the mid-transit, egress and post-transit baseline. Guiding was sub-optimal for this session due to an issue with the mount, resulting in elongation of the point spread function (PSF) in these images and an increase in photometric scatter.

\subsubsection{TIC 304130406 - TOI-3278 / HATS-78}
TIC 304130406 is a V $\approx$ 14.2 mag star in the constellation of Pavo. It was observed by \textit{TESS} in Sectors 13, 27, 67 and 94 with cadences of 1800 seconds to 200 seconds. TOI-3278.01 was announced by the Faint Star Search, with a candidate signal at 3.25 days and a depth of 8.4 ppt. We also used our custom TGLC extracted light curves for this star. This target was also identified as a planet candidate by the HATSouth survey \citep{hat-south} before the launch of \textit{TESS}, with the discovery light curve shown in the Appendix (Figure \ref{fig:HATSLCTOI-3278b}).
\newline\newline
TOI-3278 was first observed with the H50 on the night of 2024/10/09. We obtained 80 images at 120 second exposures in the SDSS $r^{\prime}$ filter, covering a small amount of pre-transit baseline, ingress and the start of egress. Conditions were clear with low moon illumination. However, observations were hindered by moderate to strong winds in the second half of the night. The dome was closed early once the safety limit (25 km~h$^{-1}$) was reached, resulting in the post-transit baseline being missed. We then reobserved TOI-3278 on 2025/10/27 in the Bessell V filter, covering similar transit phases. However, the airmass was unfavourable (starting at $\Chi \, \approx$ 2.1), resulting in a relatively poor SNR light curve. 
\newline\newline
In addition to these light curves, we obtained data from the Las Cumbres Observatory \citep[LCO,][]{lco} telescope network. Two follow-up light curves were obtained by the HATSouth survey on the 1 m telescope at Siding Spring Observatory, Australia (SSO) in 2015. Another light curve was obtained on 2021/06/13 with the 1 m telescope at the Cerro Tololo Inter-American Observatory, Chile (CTIO) through the SG1 program. The images were calibrated with the BANZAI pipeline \citep{banzai}. The CTIO differential photometry was extracted with AstroImageJ using 3.1 arcsec apertures. All LCO data was acquired in the SDSS i$^\prime$ filter, using the $4096\times4096$ SINISTRO camera (0.39 arcsec~pix $^{-1}$) or the $4000\times4000$ SBIG camera (0.23 arcsec~pix $^{-1}$).

\begin{table}[h!]
	\begin{center}
		\begin{tabular}{|l|p{1.2cm}|p{0.5cm}|p{0.5cm}|p{0.75cm}|p{0.6cm}|p{1.15cm}|}
            \toprule
            \headrow \textbf{ID} &\textbf{Dates [UTC]} & \textbf{Filter} & \textbf{Exp. Time [s]} & \textbf{Median FWHM ["]} & \textbf{N$_{Images}$} & \textbf{Airmass} \\
			\midrule 
			\textbf{TOI-3053} & 2024/12/15 & r' & 120 & 2.6 &  84 & 2.06 - 1.32 \\
			 & 2025/02/22 \newline 2025/03/04 & r' \newline r' & 150 \newline 120 & 3.2 \newline 3.0 & 37 \newline 61 & 1.16 - 1.09 \newline 1.14 - 1.08 \\
             & 2025/11/19 & g' & 90 & 3.4 & 110 & 1.83 - 1.26 \\
             & 2025/11/22 & V & 120 & 2.2 &  85 & 1.84 - 1.26 \\
             & 2025/12/03 & i' & 90 & 3.4 & 120 & 2.12 - 1.34 \\
             \midrule 
             \textbf{TOI-3272} & 2024/10/02 & r' & 120 & 2.2 & 75 & 1.37 - 1.42 \\
              & 2025/03/24 & V & 120 & 2.0 & 84 & 1.64 - 1.59 \\
              & 2025/08/19 & i' & 90 & 2.1 & 112 & 1.63 - 1.56 \\
            \midrule
            \textbf{TOI-3278} & 2024/10/09 \newline 2025/10/27 & r' \newline V & 120 \newline 150 & 3.1 \newline 2.4 & 80 \newline 70  & 1.10 - 1.30 \newline 2.10 - 1.30\\
            
			\bottomrule 
		\end{tabular}
		\caption{Observation logs for TASSIE II TOIs using the H50 telescope.}
		\label{table:tassielog}
	\end{center}
\end{table}

\subsection{Reconnaissance Spectroscopy}
To help constrain the stellar atmospheric parameters and attempt to rule out false positive scenarios, such as double-lined eclipsing binaries (EBs), we gathered low to mid resolution spectra for each target. We estimated preliminary spectral types and metallicity values using the PyHammer v2 package \citep{pyhammerv2}. Spectral indices and colour regions of input spectra were compared to SDSS templates through a $\chi^2$ minimisation. This code can also identify some double-lined spectroscopic binaries automatically.  The log of the spectroscopic observations is shown in Table \ref{table:tassiespec} and described in further detail here.

\subsubsection{Las Cumbres Observatory -  Floyds Spectrograph}
We obtained two low-resolution (R $\approx$ 500) spectra from the Floyds spectrograph on the LCO 2 m telescope at SSO for TOI-3278. The spectra cover wavelength ranges from 320-1000 nm in two orders. We bracketed our science spectra with wire and lamp frames for calibration. Data were reduced by the automatic LCO Floyds pipeline\footnote{\url{https://github.com/LCOGT/floyds_pipeline}}, which performs order rectification, flat-fielding and flux calibration. We noticed significant fringing in the infrared ($\lambda$ > 750 nm), so therefore restricted our analysis to the visible part of the spectra. We median combined the two spectra to improve the SNR. We found a mid G-dwarf template star with solar metallicity matches the data well, with no evidence for double lines seen. 

\subsubsection{ANU 2.3 m - Wide Field Spectrograph (WiFeS)}
We obtained mid-resolution (R = 7000) spectra with the Wide Field Spectrograph (WiFeS) on the Australian National University 2.3 m telescope at SSO \citep{wifes} for TOI-3053 and TOI-3272. Exposure times ranged between 180 and 300 s depending upon the target, with the aim of obtaining SNR $\geq$ 30. Ne-Ar arc frames were taken following each observation to ensure an accurate wavelength calibration. Bias, darks, flats and spectrophotometric standard frames were also collected at the start or end of each night. We utilised the R7000 grating with R560 dichroic, giving a wavelength range of 550-700 nm and velocity resolution of $\approx$ 45 km~s$^{-1}$. Data were reduced using the PyWiFeS package \citep{pywifes}. Once again, we classified these stars as mid-to-late G-dwarfs.
\newline\newline
We also extracted low precision radial velocities (RVs) from the WiFeS data using the iSpec package \citep{ispec}. Each science and standard star spectrum was cross-correlated with an atomic line mask, with the RVs extracted by fitting a 2nd-order polynomial + Gaussian to the peak of cross-correlation function (CCF). The error is calculated from the combination of the target and standard star errors to attempt to account for any zero point drifts. We achieve a precision of $\approx$4 ~km~s$^{-1}$ with these data. We ruled out any radial velocity signal greater than 12 ~km~s$^{-1}$, making it unlikely that these targets could be unblended eclipsing binaries. 

\begin{table}[h!]
	\begin{center}
		\begin{tabular}{|l|p{1.6cm}|p{1.5cm}|p{0.6cm}|p{0.6cm}|p{0.75cm}|}
			\toprule
			\headrow \textbf{ID} & \textbf{Instrument} &\textbf{Dates [UTC]} & \textbf{Exp. Time [s]} & \textbf{N$_{Spectra}$} & \textbf{Median SNR} \\
			\midrule 
			\textbf{TOI-3053} & ANU 2.3 m - WiFeS \newline($R = 7000$) & 2025/02/15 \newline 2025/02/19
            \newline 2025/03/19 & 180 &  6 &  36 \\
            \midrule 
            \textbf{TOI-3272} & ANU 2.3 m - WiFeS \newline($R = 7000$) & 2025/02/15 \newline 2025/02/19 \newline 2025/04/06 & 300 & 6 & 38 \\
            \midrule
            \textbf{TOI-3278} & LCO 2 m - FLOYDS (SSO) ($R = 500$) & 2025/07/18 & 300 & 2 & 29 \\
			\bottomrule 
		\end{tabular}
		\caption{Reconnaissance spectroscopy observation logs for TASSIE II TOIs}
		\label{table:tassiespec}
	\end{center}
\end{table}

\subsection{High Angular Resolution Imaging}
\begin{figure}
    \centering
    \includegraphics[width=1\linewidth]{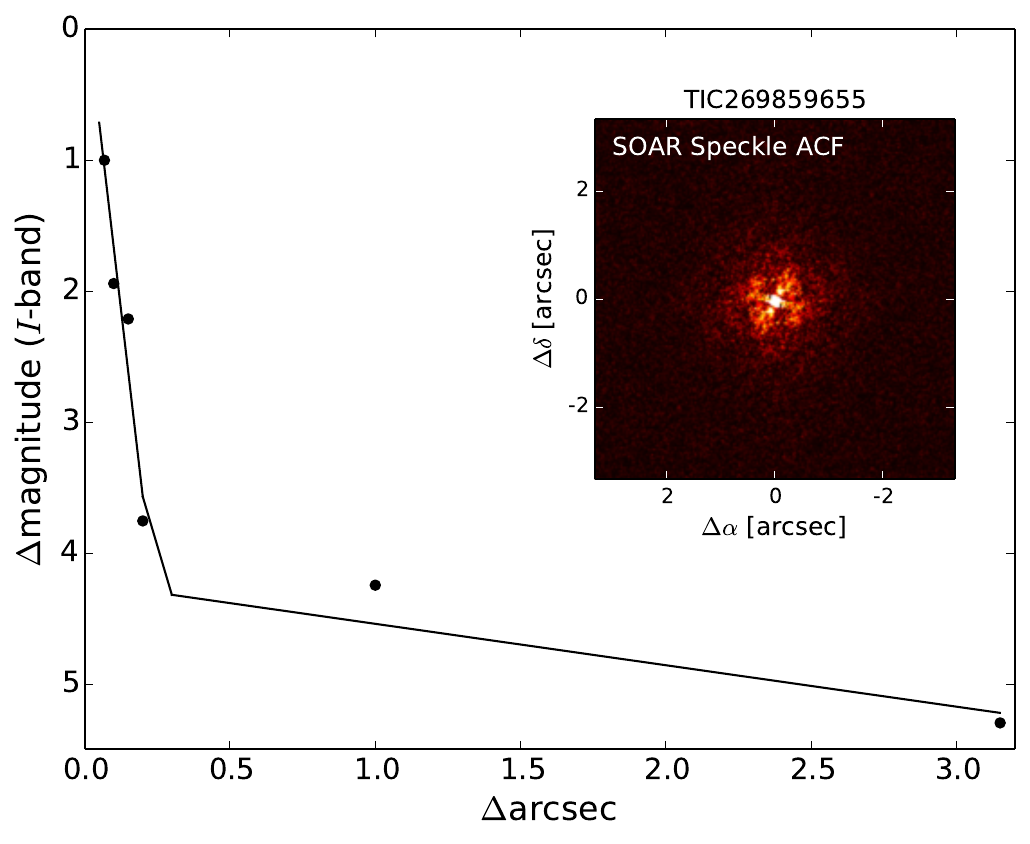}
    \caption{Contrast curve and high-resolution reconstructed autocorrelation function image for TIC269859655 (TOI-3053) from the SOAR telescope with the I filter.}
    \label{fig:speckleTOI-3053}
\end{figure}

To search for close companions, we obtained data from the HRCam Speckle Imager on the Southern Astrophysical Research (SOAR) 4.1 m telescope in I-band \citep{speckle} for TOI-3053. This is important to rule out false positives and to correct for any flux dilution of transits, which can lead to underestimation of the planetary radius \citep{ciardi2015}. For details of the data acquisition and reduction, we refer the reader to \citet{tesssoar}. The derived contrast curve and reconstructed auto-correlation function image are shown in Figure \ref{fig:speckleTOI-3053}. We identified no companions to TIC269859655 (TOI-3053) at magnitude differences of < 4 at 0.25" and < 5 out to 3". Note, we were unable to obtain high angular resolution imaging for the other two targets. 

\subsection{Radial Velocities}
\subsubsection{CHIRON}
We observed TOI-3053 with the CHIRON spectrograph on the 1.5 m telescope operated by the Small and Moderate Aperture Research Telescope System (SMARTS) consortium at CTIO \citep{chiron}. We utilised the image slicer mode with a spectral resolution of R = 79000, obtaining 11 observations with exposure times of 3600 seconds and bracketed by Th-Ar calibration spectra. The RVs were extracted via least-squares deconvolution of the observed spectra against non-rotating synthetic templates using the ATLAS9 models \citep{CastelliKurucz}, before fitting a broadening kernel to the final line profile. The median uncertainty on these measurements is 38.5 m~s$^{-1}$. The RVs are reported in the Appendix (Table \ref{table:rvlogs}), with one outlier rejected due to strong moonlight contamination.

\subsubsection{FEROS}
We obtained high-resolution spectra of TOI-3278 with the Fiber-fed Extended Range Optical Spectrograph \citep[FEROS,][]{feros} on the MPG/ESO 2.2 m at La Silla Observatory, Chile. FEROS has a spectral resolution of R = 48000 with data split over 39 echelle orders. The radial velocities were extracted using the CERES pipeline with a G2 stellar mask, as described in \citet{ceres}. A total of 34 observations were obtained with exposures varying between 1500 to 3000 seconds. Due to the faintness of the target increasing scatter, we decided to clean the data by rejecting 5$\sigma$ outliers based on the RVs. This left 24 measurements for modelling, as reported in Table \ref{table:rvlogs}. The median uncertainty on these data is 16.5 m~s$^{-1}$.

\subsubsection{HARPS}
We also obtained 3 measurements of TOI-3278 with the
High Accuracy Radial velocity Planet Searcher \citep[HARPS,][]{harps} spectrograph on the ESO 3.6 m at La Silla Observatory, Chile. HARPS possesses a spectral resolution of R = 120000, with data split over 72 echelle orders. The exposure times varied between 1500 to 1800 seconds. The high-precision radial velocities were again extracted with the CERES pipeline using a G2 mask. The median RV error is 27.1 m~s$^{-1}$. 

\section{Stellar Characterisation}
\subsection{Atmospheric Parameters}
\begin{figure*}[h!]
    \begin{subfigure}[]
	   \centering
	   \includegraphics[width = 0.45\linewidth, height = 0.3\linewidth]{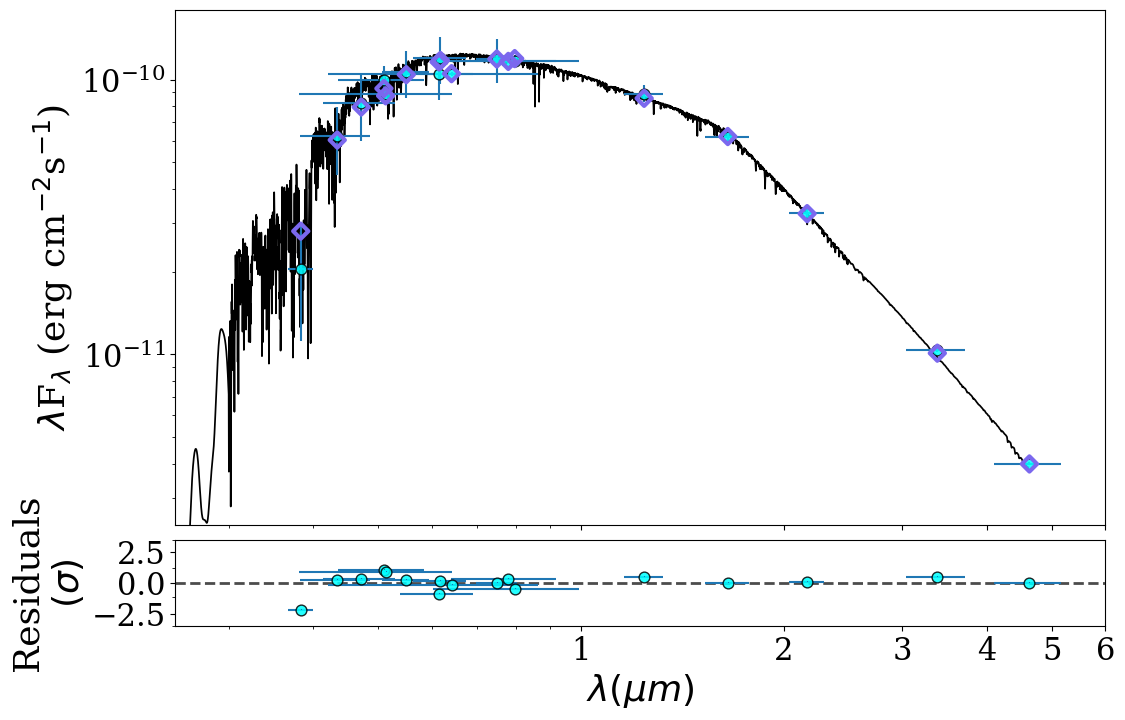}
    \end{subfigure}
    ~
    \begin{subfigure}[]
	   \centering
	   \includegraphics[width = 0.45\linewidth, height = 0.3\linewidth]{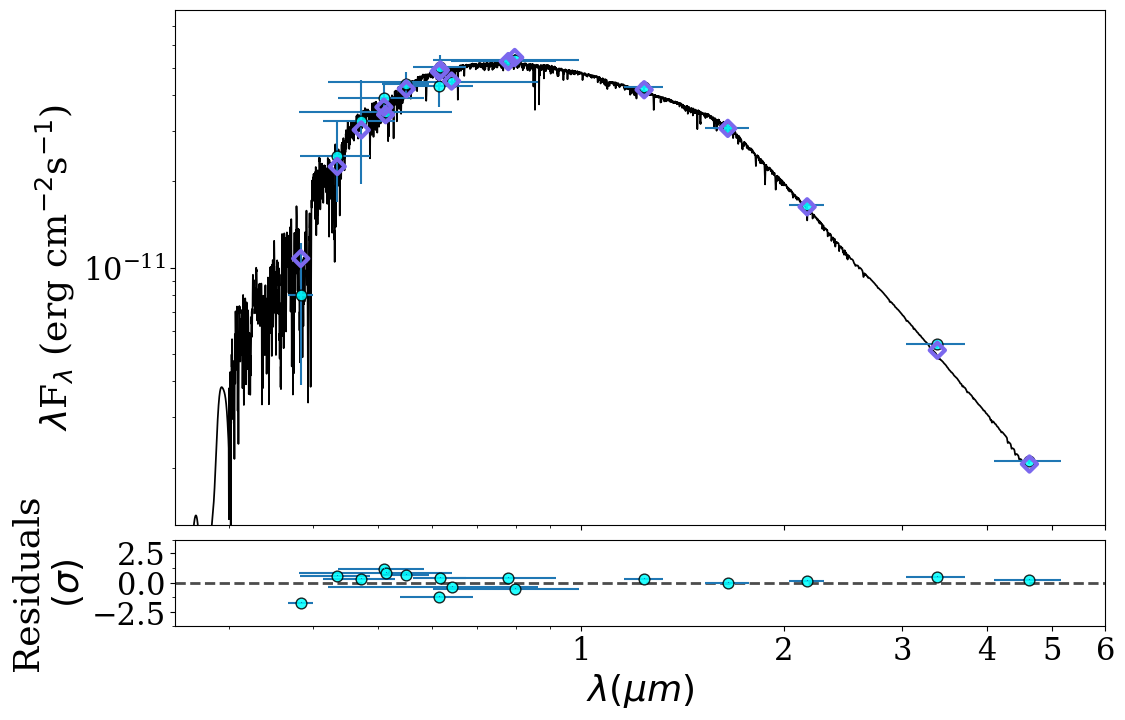}
    \end{subfigure}
    ~
    \begin{subfigure}[]
	   \centering
	   \includegraphics[width = 0.45\linewidth, height = 0.3\linewidth]{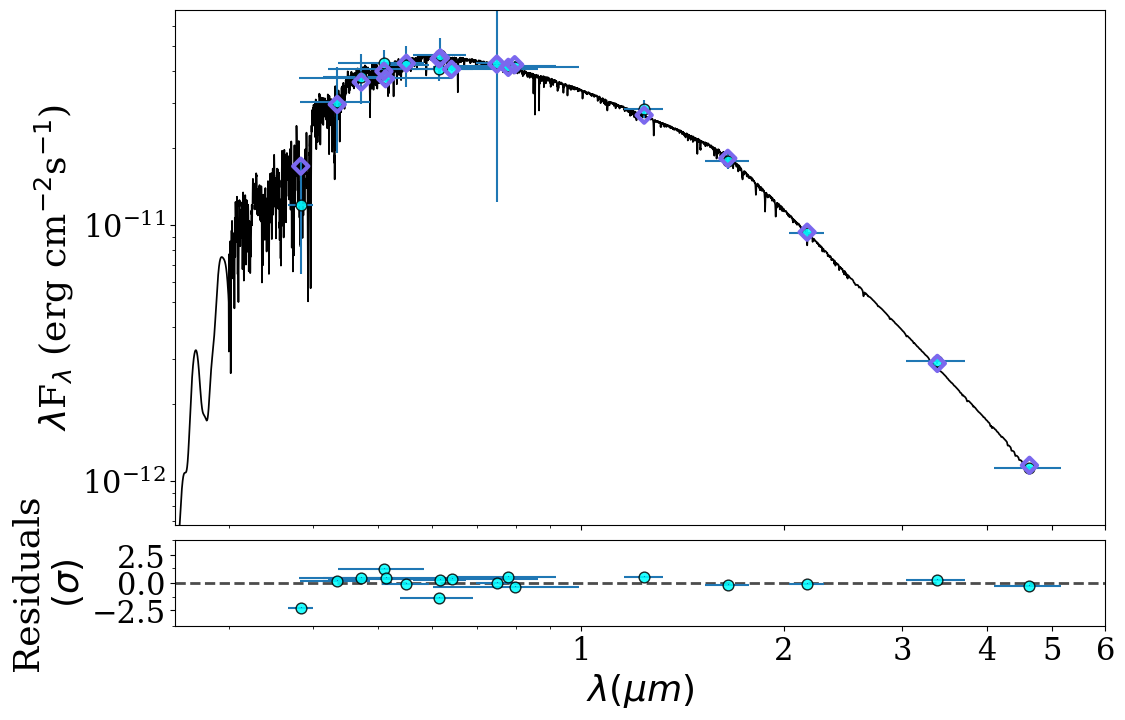}
    \end{subfigure}
	\caption{Spectral energy distributions for TOI-3053 \textbf{(a)}, TOI-3272 \textbf{(b)} and TOI-3278 \textbf{(c)} from ARIADNE. The best-fit Phoenix v2 models for each star are shown in black, along with the broadband photometry (blue points) and expected flux in each filter (purple diamonds). The residuals are shown in the bottom sub-panel normalised by the uncertainties on each point (in units of standard errors). The horizontal error bars in wavelength represent the bandpass of each filter.}
	\label{fig:SEDs}
\end{figure*}

We performed an iterative, multi-step procedure to estimate the stellar atmospheric parameters for each TOI. We started with a spectroscopic analysis using iSpec with the WiFeS spectra. After continuum normalisation (using 2nd order splines) and shifting to rest wavelengths, we median combined the individual spectra to improve the SNR. We selected various metal lines (Fe, Mg, Al, Si, Ca and Cr), fitting the line centres and profiles using Gaussian functions. We fit for four free parameters ($T_{\mathrm{eff}}$, [Fe/H], log(g), $v_{\mathrm{mic}}$) using a grid of pre-computed MARCS model atmospheres \citep{marcs}. We initially fixed $v\sin(i)$ = 2.0 ~km~s$^{-1}$ and $v_{mac}$ to values from an inbuilt empirical relation, as these parameters are degenerate and not well constrained with the resolution of the WiFeS spectra. For TOI-3278, we were unable to obtain WiFeS spectra and therefore adopted the initial estimates on these parameters from \gaia DR3 XP spectra \citep[as described in][]{tassie}.
\newline\newline 
The spectroscopic estimates were then used as normal priors for spectral energy distribution (SED) fitting with ARIADNE \citep{ariadne}. We collected all available broadband photometric data from surveys such as 2MASS \citep{2mass}, SkyMapper \citep{skymapper}, \gaia DR3 and WISE \citep{WISE}. We ran nested sampling with four model atmosphere grids, including Phoenix v2 \citep{PhoenixV2}, Kurucz 1993, ATLAS9 \citep{CastelliKurucz} and BT-Settl \citep{BTmodels}. Two extinction priors were tested, with one using the line-of-sight upper limit from the dust map of \citet{sfddustmap} (uniform distribution) and the other adopting the \gaia DR3 extinction estimate with a 10 $\%$ width (normal distribution). Whilst both priors produced consistent values, we decided to use the \gaia DR3 based runs as these led to slightly better fits to the data. The final results were derived from Bayesian model averaging of the individual posterior samples, with the uncertainties estimated from the highest density region at the 68 \% confidence level. The SEDs are shown in Figure \ref{fig:SEDs} and the derived atmospheric parameters are listed in Table \ref{table:stellparams}. Finally, we attempted to constrain the projected rotational velocity by once again using iSpec. We fixed all other parameters to their previously derived values and fitted broadening kernels to synthetic spectra to match the median combined WiFeS or FEROS data. We were only able to constrain that $v\sin(i)$ < 20 ~km~s$^{-1}$ for TOI-3053 and TOI-3272. For TOI-3278, we found a tighter constraint of $v\sin(i)$ < 2.0 ~km~s$^{-1}$.

\subsection{Stellar Radius and Mass}
Using the \gaia DR3 parallax as a constraint, ARIADNE estimates the distance and radius of each star by scaling the stellar model fluxes to the observed fluxes. We inferred radii of $R_{3053}$ = 0.94 $\pm$ 0.01 $R_{\odot}$, $R_{3272}$ = 1.27 $\pm$ 0.03 $R_{\odot}$ and $R_{3278}$ = 1.39 $\pm$ 0.03 $R_{\odot}$ for TOI-3053, TOI-3272 and TOI-3278, respectively. We then interpolated with MESA isochrones and Stellar Tracks \citep[MIST, ][]{mist} models to estimate the masses, which are summarised in Table \ref{table:stellparams}. As an independent check of these values, we used the empirical relations of \citet{Torres2010}. These relations were derived from the measured masses and radii of well-detached eclipsing binaries. Using our estimates of $T_{eff}$, log($g$) and $[Fe/H]$, we found consistent values for both radius and mass (with larger uncertainties). We adopted the radius and mass estimates from the SED fitting  due to the higher precision and consistency with other values. 

\subsection{Kinematics}
We calculated the galactic velocities of each TOI using the equations defined by \citet{soderblom1987}, with an updated transformation matrix for the J2000 epoch. We took the parallax, proper motions and radial velocities in Table \ref{table:stellparams} from \gaia DR3 \citep{dr3}. We adopted the solar velocities of ($U_{\odot}, V_{\odot}, W_{\odot}$) = (8.5, 13.38, 6.49) ~km~s$^{-1}$ from \citet{cos2011}. Here, U is defined as positive towards the centre of the galaxy, V is positive in the direction of galactic rotation and W is positive in the direction of the North galactic pole. The results of this analysis are listed in Table \ref{table:stellparams}. Using the membership probabilities defined by \citet{bensby2003}, we placed all targets in the thin disk. We also utilised the BANYAN $\Sigma$ tool \citep{banyan} to check if any of the TOIs are members of known young stellar associations. We found these are likely field stars with 99.9\% probability. 

\subsection{Stellar Activity, Rotation Periods and Ages}
\begin{figure}[h!]
    \centering
    \includegraphics[width=0.99\linewidth, height = 0.9\linewidth]{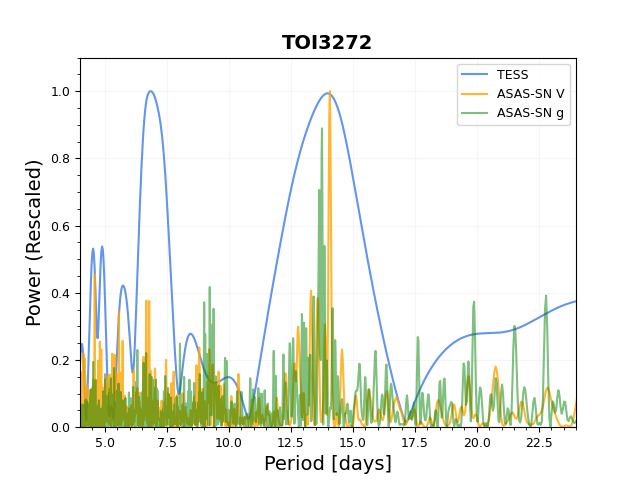}
    \caption{Lomb-Scargle periodogram for TIC 388280249 (TOI-3272). The blue curve shows the \textit{TESS} periods, with the green and orange corresponding to ASAS-SN g and V band periods, respectively. Note, the power for each has been rescaled by dividing by the maximum power value for each dataset.}
    \label{fig:TOI-3272_ls}
\end{figure}

Quasi-periodic variability is seen in the light curve of TOI-3272 at the 1-2 \% level, consistent with the signatures of star spots. To help constrain the rotational period, we investigated archival photometry from the ASAS-SN \citep{ASASSN} survey which has a longer observing span than \textit{TESS}. After cleaning erroneous measurements using sigma clipping (at 5 $\sigma$ and 10 iterations), we performed a Generalised Lomb-Scargle (GLS) periodogram search \citep{gls} as shown in Figure \ref{fig:TOI-3272_ls}. We found two peaks at $\approx$ 6.9 and 13.9 days from \textit{TESS} data. We determined a similar value of 14.09 days in the ASAS-SN V data, with an alias of 29.95 days in ASAS-SN $g^{\prime}$. As a further step in determining which was the true period, we modelled the \textit{TESS} data with a Gaussian Process through the Celerite package \citep{celerite}. We choose a quasi-periodic kernel and provide a broad uniform prior between 5 to 20 days. We recover a consistent estimate of $P_{rot}$ = $13.44 \pm 0.46$ days and adopt this as the rotational period of TOI-3272. Using the gyrochronology relations of \citet{hillenbrand2008} with the known colour of the star, this would imply an age of $1.1 \pm 0.2$ Gyr. This places TOI-3272 as a young to intermediate age field star. For TOI-3053 and TOI-3278, we see no clear signs of variability in the \textit{TESS} or ASAS-SN light curves. This is validated through a GLS search of each dataset returning no matching periods of significance. We also ruled out any periodic signals and RV correlations in the BIS measurements from the FEROS timeseries for TOI-3278. Without rotation periods, we can only place loose age constraints from MIST isochrone fitting (which can have large systematic errors for main-sequence stars). We estimated ages of $7.8^{+1.3}_{-6.2}$ Gyr and $8.9^{+2.9}_{-2.3}$ Gyr for TOI-3053 and TOI-3278 / HATS-78, respectively. 

\begin{table*}[h!]
	\begin{center}
		\begin{tabular}{|p{2.5cm}|p{2.5cm}|p{2.5cm}|p{2.5cm}|p{3.5cm}|} 
			\toprule
			\headrow \textbf{Parameters} & 
            \textbf{TOI-3053} & \textbf{TOI-3272} & \textbf{TOI-3278 / HATS-78} & \textbf{Source} \\
			\midrule 
			\textbf{Catalogue IDs} \newline 2MASS \newline \gaia DR3 \newline UCAC4  & TIC 269859655 \newline J09541767-6417489 \newline 5249550994759610624 \newline 129-025150 	& TIC 388280249 \newline J08250176-8509284 \newline 5190248147907583104 \newline 
025-003206 & TIC 304130406 \newline J19272804-6205413 \newline 6444865991227611136 \newline 140-195552 & Simbad \\
			\hline\hline 
			\headrow \multicolumn{5}{|c|}{\textbf{Measured Properties}}\\
			\hline\hline 
            \textbf{RA} [hour angle] & 09:54:17.665 & 08:25:01.820 & 19:27:28.054 & Simbad \\
            \hline
            \textbf{Dec} [degrees] & -64:17:48.927 & -85:09:28.367 & -62:05:41.383 & Simbad \\
            \hline
            \textbf{\gaia G} [mag] & $12.961 \pm 0.003$ & $13.891 \pm 0.003 $ & 13.984 $\pm$ 0.003 &  \gaia DR3 \\
            \hline
            \textbf{\gaia BP-RP} [mag]  & $1.08 \pm 0.01$ & $1.23 \pm 0.01$ & 0.89 $\pm$ 0.01 & \gaia DR3\\
			\hline
			\textbf{V} [mag]  & $13.19 \pm 0.04$ & $14.15 \pm 0.01$ & 14.18 $\pm$ 0.03 & UCAC4 \citep{UCAC4} \\
	        \hline 
			\textbf{Ks} [mag]  & $11.13 \pm 0.02$ & $11.87 \pm 0.03$ & 12.49 $\pm$ 0.03 & 2MASS \citep{2mass} \\
			\hline
			\textbf{Parallax} [mas] & $2.9837 \pm 0.0137$ & $1.6201 \pm 0.0132$ & 1.0819 $\pm$ 0.0160 & \gaia DR3\\ 
			\hline
			\textbf{PM}$_{RA}$ \newline [mas~yr$^{-1}$] & $-38.501 \pm 0.015$ & $-12.950 \pm 0.019$ & -3.620 $\pm$ 0.015 & \gaia DR3 \\
			\hline
			\textbf{PM}$_{DEC}$ \newline [mas~yr$^{-1}$] & $4.687 \pm 0.014$ & $18.594 \pm 0.015$ & -11.227 $\pm$ 0.014 & \gaia DR3\\
            \hline 
            \textbf{RUWE} & 1.02 & 1.05 & 1.06 & \gaia DR3 \\
			\hline\hline
		    \headrow \multicolumn{5}{|c|}{\textbf{Derived Properties}} \\
		    \hline\hline 
			\textbf{Extinction} $A_{V}$ [mag] & $0.28^{+0.03}_{-0.03}$ & $0.70^{+0.05}_{-0.05}$ & $0.14^{+0.02}_{-0.02}$ & This Work\\
			\hline
			\textbf{Effective Temp.} [K] & $5378^{+82}_{-90}$  & $5526^{+69}_{-81}$ & $5720^{+89}_{-91}$ & This Work \\
			\hline
			\textbf{log($\frac{\textbf{g}}{[cm~s^{-2}]}$)}  &  $4.34^{+0.10}_{-0.10}$  & $4.36^{+0.15}_{-0.14}$ & $4.26^{+0.08}_{-0.08}$ &  This Work\\
			\hline
			\textbf{Radius} [$R_{\odot}$] &  $0.97^{+0.01}_{-0.02}$  & $1.27^{+0.03}_{-0.03}$ & $1.39^{+0.03}_{-0.03}$ & This Work \\
			\hline
			\textbf{Mass} [$M_{\odot}$] & $0.96^{+0.08}_{-0.03}$ & $0.98^{+0.08}_{-0.04}$ & $1.04^{+0.09}_{-0.10}$ &  This Work\\
			\hline
			\textbf{Distance} [pc] & $336.6^{+1.6}_{-0.9}$ & $622.6^{+4.0}_{-3.0}$ & $940.6^{+14.8}_{-8.2}$& This Work \\
			\hline
			\textbf{[Fe/H]} (dex) & $0.20^{+0.05}_{-0.05}$ & $0.17^{+0.06}_{-0.06}$ & $0.04^{+0.14}_{-0.14}$ &  This Work\\
			\hline
			\textbf{Spectral Type} & G8 - K0 V  & G6 - G8 V & G2 - G4 V & \citet{mamajek}\\
            \hline 
            \textbf{Rot. Period} [days] & - & 13.44 $\pm$ 0.46 & - & This Work \\
			\hline
            \textbf{Age} [Gyr] & $7.8^{+1.3}_{-6.2}$ & $1.1^{+0.2}_{-0.2}$ & $8.9^{+2.9}_{-2.3}$ & This work \\
            \hline 
            \textbf{v sin(i)} [km~s$^{-1}$] & < 18.7 $\pm$ 2.4 & < 18.1 $\pm$ 3.6 & < 2.0 $\pm$ 0.1 & This Work \\
			\hline
            \textbf{v$_{mic}$} [km~s$^{-1}$] & $0.6 \pm 0.2$  & $0.5 \pm 0.2$ & 1.0 $\pm$ 0.2 & This Work\\
			\hline\hline
			\headrow \multicolumn{5}{|c|}{\textbf{Kinematic Properties}}\\
			\hline\hline
			\textbf{RV} [km~s$^{-1}$] &  47.42 $\pm$ 1.01 & 47.59 $\pm$ 2.70 & 31.17 $\pm$ 3.03 & \gaia DR3\\
			\hline
			\textbf{U} [km~s$^{-1}$] &  -30.82 $\pm$ 0.34 & -31.1 $\pm$ 1.2 & 16.1 $\pm$ 2.4 & This Work\\
			\hline 
			\textbf{V} [km~s$^{-1}$] &  -41.40 $\pm$ 1.00 &  -51.4 $\pm$ 2.2 & -46.3 $\pm$ 1.4 & This Work\\
			\hline
			\textbf{W} [km~s$^{-1}$] &  -32.12 $\pm$ 0.19 & -23.4 $\pm$ 1.1  & 1.4 $\pm$ 1.4 & This Work \\
            \bottomrule 
		\end{tabular}
		\caption{Stellar parameters for TASSIE II TOIs. The reported errors on SED derived parameters account for both statistical and model uncertainties.}
		\label{table:stellparams}
	\end{center}
\end{table*}

\section{Data Analysis and Modelling}
\subsection{Planet Validation}
\subsubsection{Astrometry}
One method for checking for multiplicity in stellar systems is through the \gaia DR3 astrometric solution. In particular, we can look at the Renormalized Unit Weight Error (RUWE), which is a weighted chi-square test of the astrometry (accounting for the colour and magnitude of the target). \gaia reports RUWEs of $\approx$ 1 for each of these targets, which is well below the threshold of 1.4 that usually hints towards unresolved companions \citep{RUWE}. This helps to rule out eclipsing binaries with scales of 0.1-10 AU as the source of the transit signals.

\subsubsection{Archival Imaging}
To check for background EB scenarios, we investigated archival imaging data from the Harvard College Observatory's Photographic Glass Plate Collection using the Dasch package \citep{Dasch}. We searched data dating back to the late 1890's for the best image quality and limiting magnitude. We selected frames from between 1930-1955 with exposures of 2700 s. By comparing catalogues of sources extracted from stacked H50 images for each target with these earlier epochs, we found no obvious background stars with B < 17 mag. This reduces the likelihood that the transit signals could be caused by background EBs. However, if the relative proper motion of the target and a background star were low enough (below $\approx$ 10 mas~yr$^{-1}$), then it is possible that a background EB could remain blended even after 70 years. The low proper motion of these target stars therefore makes it difficult to completely rule out such a scenario.


\subsubsection{Transit Depth Chromaticity}
\begin{figure}[h!]
    \begin{subfigure}[]
	   \centering
	   \includegraphics[width = 0.99\linewidth, height = 0.6\linewidth]{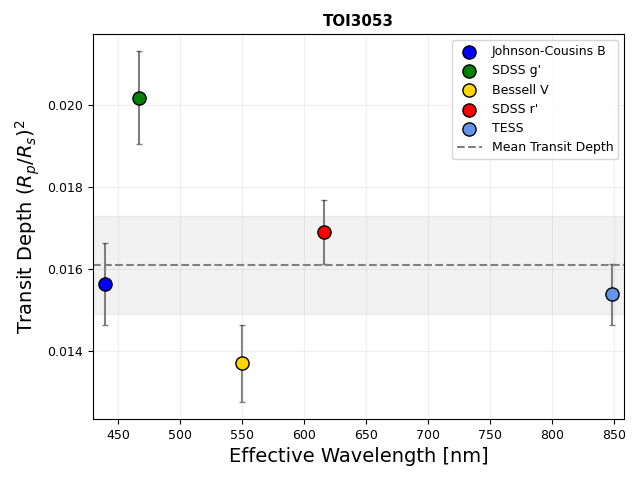}
    \end{subfigure}
    ~
    \begin{subfigure}[]
	   \centering
	   \includegraphics[width = 0.99\linewidth, height = 0.6\linewidth]{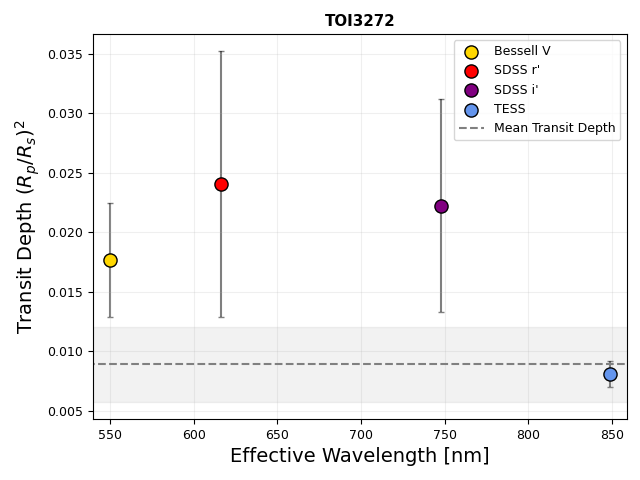}
    \end{subfigure}
    ~
    \begin{subfigure}[]
	   \centering
	   \includegraphics[width = 0.99\linewidth, height = 0.6\linewidth]{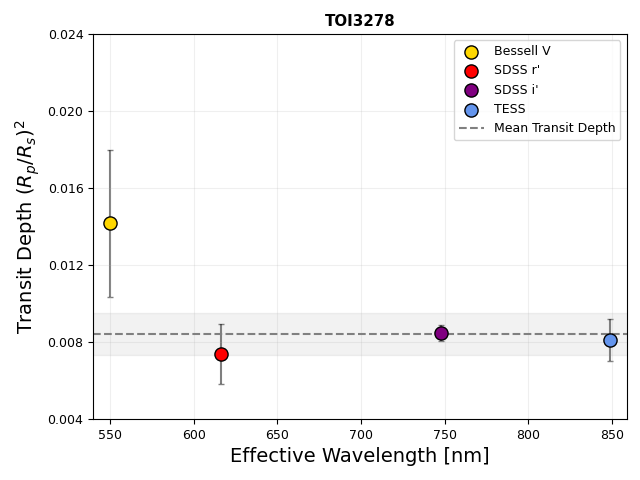}
    \end{subfigure}
	\caption{Transit depth as a function of effective wavelength for TOI-3053 \textbf{(a)}, TOI-3272 \textbf{(b)} and TOI-3278 \textbf{(c)} from individual fitting with Juliet. The error bars indicate the $1\sigma$ range, with the grey dashed line and shaded regions indicating the weighted mean depth and 99.8 $\%$ confidence interval. The transit appears achromatic for TOI-3053 and TOI-3278, with all points lying within $3\sigma$ of the mean (and the \textit{TESS} value). However, TOI-3272 remains ambiguous due to the large uncertainties.}
	\label{fig:chromaticity}
\end{figure}

Another method to rule out false positives is to check that the transit depths across different wavelength regions are consistent. We expect planetary transits to be achromatic in broadband filters, with deviation from this condition indicating the signal could be a result of an EB system instead. To check for chromaticity in the transits, we used Juliet \citep{juliet} to fit each light curve from \textit{TESS} and the ground-based telescopes individually. We applied strict priors on the orbital period, transit epoch and limb darkening coefficients, but allowed the transit depth and impact parameter to explore the full planetary parameter space (with an upper limit of $R_{p}/R_{s}$ < 0.2). We then extracted the transit depth (by squaring the radius ratio) and compared these against the effective wavelength of each filter used. The results are summarised in Figure \ref{fig:chromaticity}. 
\newline\newline
We can see that TOI-3053.01 and TOI-3278.01 are achromatic, with the transit depths across different filters agreeing within $3 \, \sigma$ of the weighted mean (and the \textit{TESS} value). The minor deviation observed for the SDSS $g^{\prime}$ filter with TOI-3053.01 is likely due to poorer data quality from inclement weather. However, the results for TOI-3272.01 remain ambiguous. The depths do not show tight agreement between \textit{TESS} and the ground-based observations (noting that the wide uncertainties on the H50 measurements means they are statistically consistent). There are two explanations for this. Firstly, as the observations with the H50 do not cover the full transit and have limited baseline coverage, we cannot model for the variability of the star accurately. The spot coverage is changing with each epoch, which is known to alter the measured transit depth if unaccounted for \citep[e.g.,][]{beaulieu2008}. Spot modulation should affect shorter wavelength filters more, due to the higher contrast between the cooler starspots and surface in this spectral region. The second possibility is that this is a blended or grazing eclipsing binary system. Given the data, it is difficult to ascertain which scenario is most likely from this test alone.

\subsubsection{Statistical Vetting with TRICERATOPS}
To further test for false positive scenarios, we used TRICERATOPS \citep{triceratops} with the high cadence (200 s) \textit{TESS} data and previously derived stellar parameters. The code calculates the false positive probability (FPP) and nearby false positive probability (NFPP) by fitting models for EBs (primary, diluted, background and nearby) and for transits around a primary, secondary or background star. We removed many stars from consideration due to the detection of the transit signals with the H50. Where available, we included the contrast curves from high angular resolution imaging to further constrain the probabilities. We ran 10 iterations of this fitting procedure to ensure consistency in results for each planet candidate. The mean and standard error are reported in Table \ref{table:triresults}. For TOI-3053.01, TRICERATOPS suggests a very low probability of this being a false positive signal. However, we note that we cannot statistically validate a giant planet with these tests alone because of the overlap in their radius distributions with brown dwarfs and very low mass stars. The FPPs for the other targets suggest that these signals are also likely planetary in nature, but with less certainty. We therefore proceeded to model all the targets further in a global fit.

\begin{table}[h!]
	\begin{center}
		\begin{tabular}{|l|p{2cm}|l|p{2.5cm}|}
			\toprule
			\headrow \textbf{ID} &\textbf{FPP} & \textbf{NFPP} & \textbf{Most Likely Scenario} \\
			\midrule 
			\textbf{TOI-3053} & (3$\pm$0.3) x$10^{-4}$
            \newline\newline 
            [ (1$\pm$0.3) x$10^{-4}$ ]
            & $\leq$ 0.0001 & Transiting Planet (TP) \\
            \midrule
			\textbf{TOI-3272} & 0.38 $\pm$ 0.02 & $\leq$ 0.0001 & TP \\
            \midrule
            \textbf{TOI-3278} & 0.30 $\pm$ 0.02 & $\leq$ 0.0001 & TP \\
			\bottomrule 
		\end{tabular}
		\caption{Vetting results from TRICERATOPS for TASSIE II TOIs. Values in square brackets include the constraint from high angular resolution imaging.}
		\label{table:triresults}
	\end{center}
\end{table}

\subsection{Global Modelling with Juliet}
We modelled each planetary system with the Juliet package \citep{juliet}, which uses the BATMAN transit and RadVel RV fitting codes \citep{batman, radvel}. We explore the parameter space via nested sampling as implemented in Dynesty \citep{dynesty}, ensuring that the number of live points exceeds the number of parameters squared ($i.e., n_{points} > n_{param}^2$). In our modelling, the limb darkening coefficients are re-parameterised following \citet{kipping2013}, with normal priors set with estimates from the ExoCTK limb darkening calculator\footnote{\url{https://exoctk.stsci.edu/limb_darkening}} using Phoenix v2 stellar models. The dilution factor for the \textit{TESS} light curves is set to 1 (no dilution), as they are already de-blended. It was found that only TOI-3272 had non-negligible dilution in the ground-based light curves from a close non-companion star, which we accounted for with a uniform prior with ranges based upon flux ratio estimates from the TRICERATOPS vetting process. We chose to treat each group of cadences from \textit{TESS} as an individual instrument (i.e., 600 s cadence data are grouped together). This was done to account for different noise levels and systematics from different cadences. The full description of the parameters and priors used in these modelling runs are shown in the Appendix (Table \ref{table:priors}). We now detail each system individually, with all parameter estimates listed in Table \ref{table:planetresults} and posterior distributions shown in Figure \ref{fig:posteriors}.

\subsubsection{TOI-3053}
\begin{figure*}[h!]
    \begin{subfigure}[]
    \centering
    \includegraphics[width=0.99\linewidth, height=0.3\linewidth]{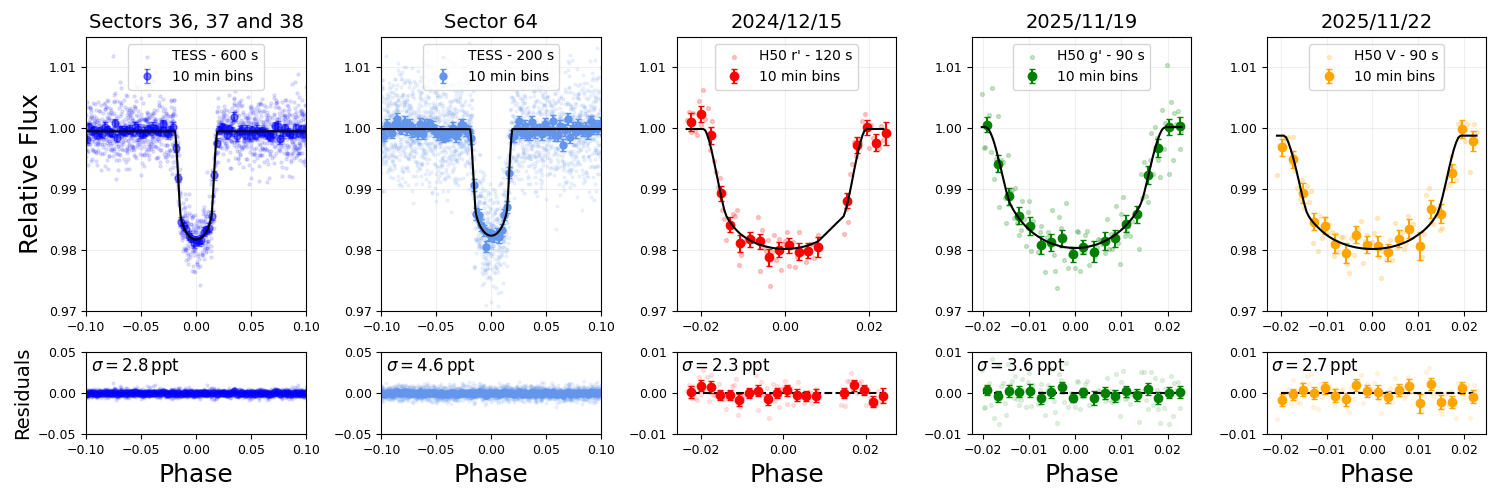}   
    \end{subfigure}
    ~
    \begin{subfigure}[]
    \centering
    \includegraphics[width=1\linewidth, height=0.35\linewidth]{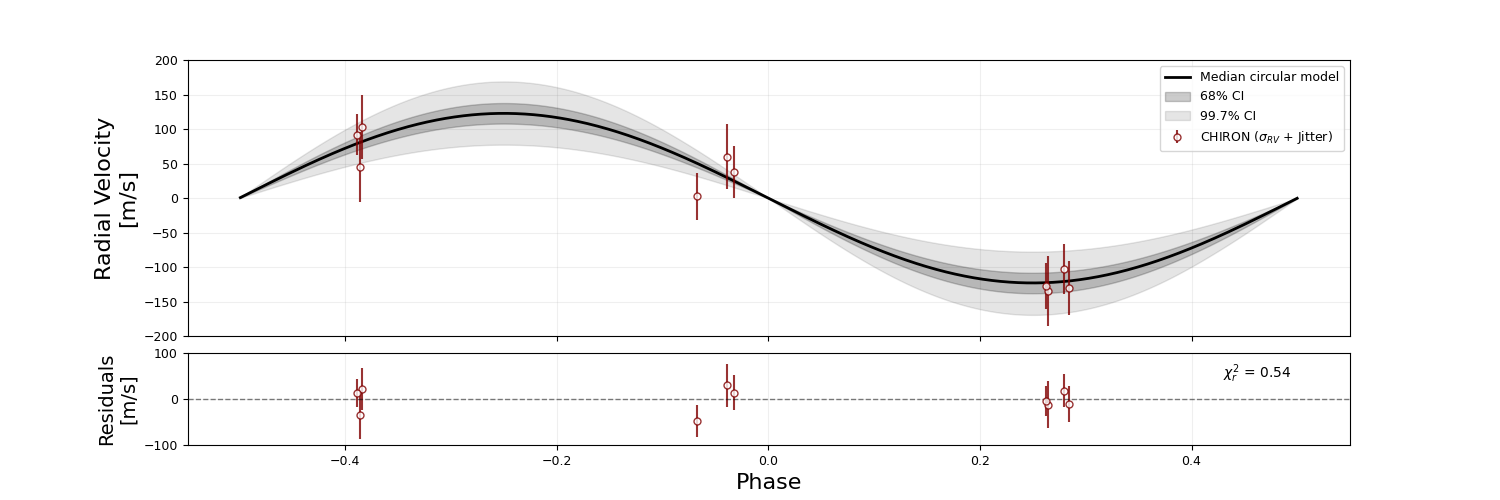}   
    \end{subfigure}
    \caption{The phase-folded light curves and RVs for TOI-3053b for each instrument/cadence. In \textbf{(a)}, the best-fit transit model is displayed after the subtraction of the GP model for each instrument. The coloured points with error bars represent the photometric data binned to 10-minutes for clarity. The panels below each light curve show the residuals and measured photometric scatter. In \textbf{(b)}, the CHIRON RVs (maroon points) and the best-fit circular model (solid black line) are shown. The light and dark shaded region shows the $1\sigma$ and $3\sigma$ confidence intervals, respectively.}
    \label{fig:TOI-3053_folded}
\end{figure*}

For TOI-3053, we decided to only model data from Sectors 36, 37, 38 and 64 (using the QLP extraction), plus the SDSS $r^{\prime}$, SDSS $g^{\prime}$ and Bessell V filters from the H50 and the CHIRON RVs. This was done in the interest of minimising run time while retaining the higher cadence data that had reasonable coverage of the transit profile. We experimented with multiple detrending approaches, in the end favouring two types of GPs: a quasi-periodic kernel (QPK) for \textit{TESS} and a Matern 3/2 kernel for the H50. We chose the QPK to remove the periodic nuisance signal of the nearby blended eclipsing binary (discussed in Section 2). For the QPK, we set the rotation period hyperparameter to be shared for both \textit{TESS} instruments and set a normal prior of $P_{rot}$ = $0.4 \pm 0.1$ days, based on the period expected from previous BLS periodogram analysis. The phase-folded transit and RVs for each instrument/cadence are shown in Figure \ref{fig:TOI-3053_folded}, with the additional ground-based observations and the multi-sector light curves from \textit{TESS} displayed in the Appendix (Figures \ref{fig:TOI3053SG1} and \ref{fig:TOI-3053}). We find this companion is a hot Jupiter, with $R_{3053b}$ = $1.21 \pm 0.03$ $R_{J}$, $M_{3053b}$ = $0.85 \pm 0.12$ $M_{J}$ and a period of $P_{3053b}$ = $2.9919986 \pm 0.0000012$ days. A circular orbit is preferred (e = 0 and $\omega$ = 90 deg), with the eccentric model having a lower log-likelihood and more parameters. We also ran a fit for transit timing variations (TTVs), fixing all parameters to the best-fit circular model and providing normal priors on each individual transit. We find no evidence for TTVs in this system, with the O-C graph shown in Figure \ref{fig:TTVs} (a).

\subsubsection{TOI-3272}
We modelled the data from \textit{TESS} Sectors 27, 38, 39, 65, 66, 67, 93, 94 and 95, along with all of the H50 light curves for TOI-3272. We chose to again use a QPK kernel on \textit{TESS} to account for the stellar activity signal, with a 3/2 Matern kernel for the H50. We show the detrended, phase-folded transits in Figure \ref{fig:TOI-3272_folded} and the full \textit{TESS} light curves with best-fit transit + GP model in the Appendix (Figure \ref{fig:TOI-3272}). We find a companion with $R_{3272} = 1.15 \pm 0.05$ $R_{J}$ and a period of $P_{3053b} = 3.1466961 \pm 0.0000057$ days. A circular orbit is once again preferred and no TTVs were detected. However, the stellar activity combined with the relative faintness of the target resulted in large timing uncertainties ($> 30$ minutes) in some \textit{TESS} sectors, as shown in Figure \ref{fig:TTVs} (b). It is important to also note that the posterior distributions of the radius ratio and impact parameter display no severe degeneracies or bi-modalities. This indicates that this system is likely planetary in nature, despite the initial ambiguity in transit depths from individual fitting. Nevertheless, high-precision RVs will be required to confirm the target is in the planetary-mass regime.

\begin{figure*}[h!]
    \centering
    \includegraphics[width=0.99\linewidth, height=0.4\linewidth]{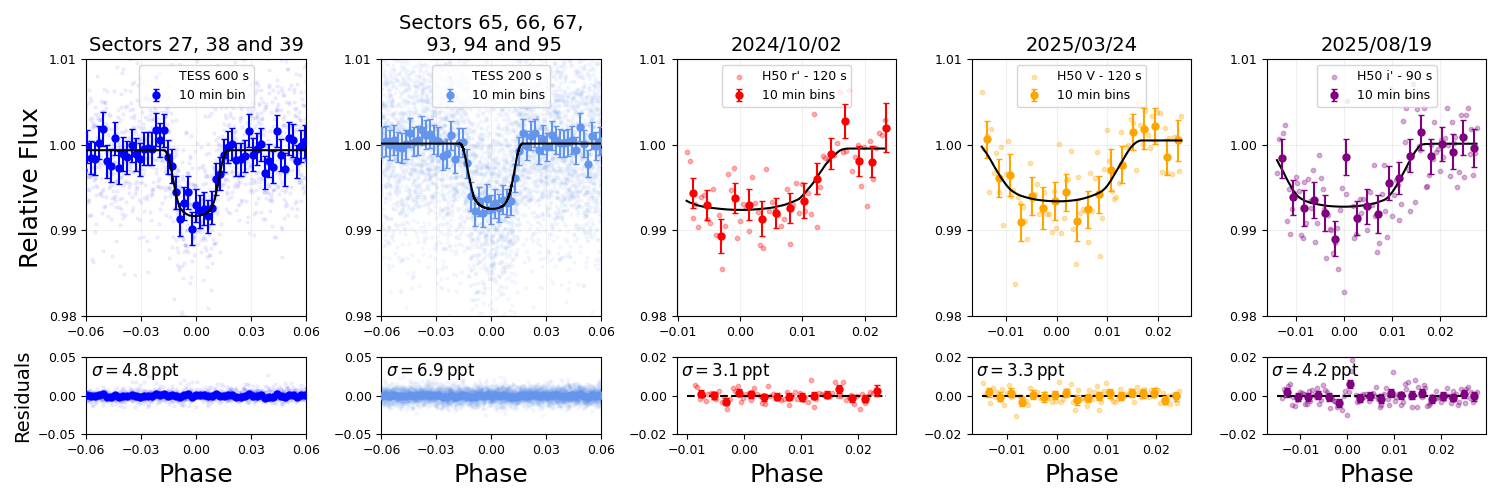}
    \caption{The phase-folded light curves for TOI-3272.01 for each cadence/instrument with best-fit transit model (in black), after the subtraction of the median GP model for each instrument. The coloured points with error bars represent the photometric data binned to 10-minutes for clarity. The panels below each light curve show the residuals and measured photometric scatter.}
    \label{fig:TOI-3272_folded}
\end{figure*}

\subsubsection{TOI-3278 / HATS-78}
\begin{figure*}[h!]
    \begin{subfigure}[]
    \centering
    \includegraphics[width=0.99\linewidth, height=0.3\linewidth]{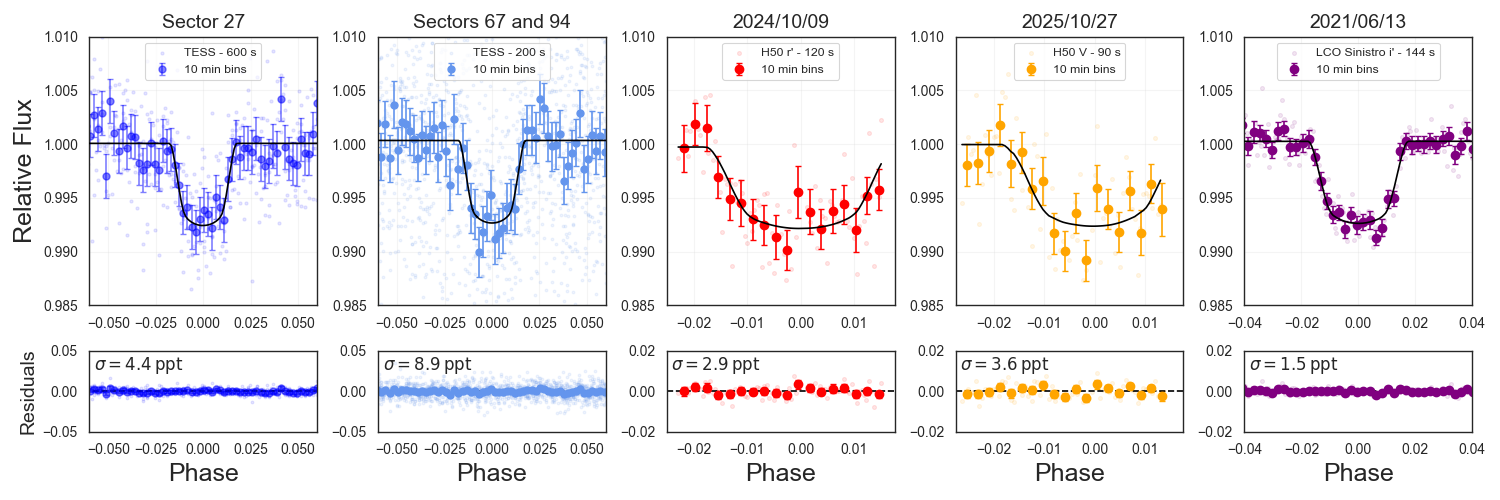}   
    \end{subfigure}
    ~
    \begin{subfigure}[]
    \centering
    \includegraphics[width=1\linewidth, height=0.35\linewidth]{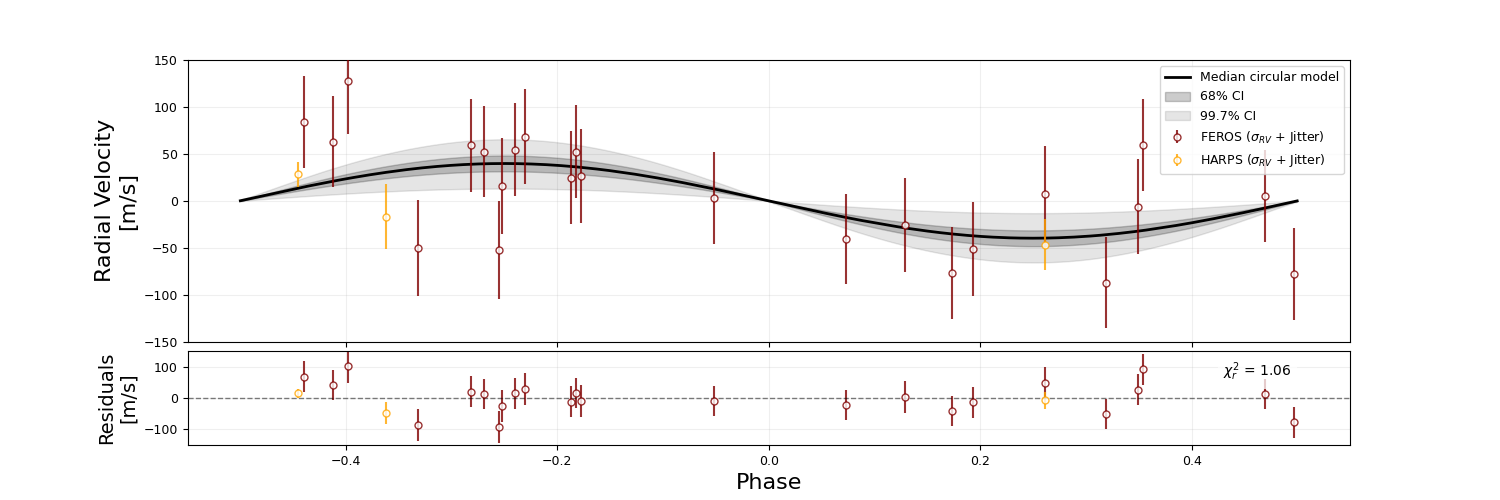}   
    \end{subfigure}
    \caption{The phase-folded light curves and RVs for TOI-3278b / HATS-78b for each instrument/cadence. In \textbf{(a)}, the best-fit transit model is displayed after the subtraction of the GP model for each instrument. The coloured points with error bars represent the photometric data binned to 10-minutes for clarity. The panels below each light curve show the residuals and measured photometric scatter. In \textbf{(b)}, the RVs are coloured by spectrograph with the best-fit circular Keplerian RV model shown by the solid black line. The light and dark shaded region shows the $1\sigma$ and $3\sigma$ confidence intervals, respectively.}
    \label{fig:TOI-3278_folded}
\end{figure*}

For TOI-3278, the final modelling run was performed with the photometry from \textit{TESS} Sectors 27, 67, 94 (using the TGLC extraction), two H50 light curves and the LCO CTIO light curve, along with the FEROS and HARPS RVs. Whilst we initially included Sector 13 as well, we discovered that the noise/systematics were too severe to provide meaningful information about the transit and we therefore discarded these data to avoid biasing the fit. We applied a Matern 3/2 GP to the \textit{TESS}, H50 and CTIO light curves, as there was no indication of periodic stellar activity from the previous periodogram analysis. We explored both circular and eccentric models, finding the circular model was preferred ($\Delta\ln Z = 2.03$). The phase-folded transits and RVs are shown in Figure \ref{fig:TOI-3278_folded}. The multi-sector \textit{TESS} light curves and RV timeseries with best-fit transit + RV model, along with the two additional LCO light curves, are shown in the Appendix (Figures \ref{fig:LCOFOLLOWUPTOI-3278b} and \ref{fig:TOI-3278}). We confirm that TOI-3278 is orbited by an inflated Saturn-mass planet, with $R_{3278b}$ = 1.24 $\pm$ 0.04 $R_{J}$, $M_{3278b}$ = 0.30 $\pm$ 0.07 $M_{J}$ and $P_{3278b}$ = 3.246456 $\pm$ 0.000005 days. Adopting the best-fit circular model parameters, we searched for TTVs and found no evidence for signals above 10 minutes. This is shown in Figure \ref{fig:TTVs} (c).

\begin{table*}[h!]
	\begin{center}
		\begin{tabular}{|l|l|l|l|} 
			\toprule
			\headrow \textbf{Parameters} & \textbf{TOI-3053b} & \textbf{TOI-3272.01} & \textbf{TOI-3278b / HATS-78b}\\
            \midrule
			\headrow \multicolumn{4}{|c|}{\textbf{Fitted Parameters}} \\
            \midrule
			$P_{b}$ [days] & $2.9919987^{+0.0000013}_{-0.0000012}$ & $3.1466961^{+0.0000057}_{-0.0000058}$ & $3.246456^{+0.0000045}_{-0.0000044}$ \\ 
            $T_{0}$ [BJD-2457000] & $2358.50739^{+0.00033}_{-0.00033}$ & $2387.0599^{+0.0019}_{-0.0019}$ & $1680.8134^{+0.0015}_{-0.0015}$ \\ 
            $R_{p}/R_{s}$ & $0.1250^{+0.0017}_{-0.0016}$ & $0.0905^{+0.0031}_{-0.0031}$ & $0.0897^{+0.0018}_{-0.0019}$ \\ 
            $b$ & $0.404^{+0.051}_{-0.070}$ & $0.822^{+0.019}_{-0.020}$ & $0.814^{+0.018}_{-0.019}$ \\ 
            $a/R_{s}$ & $8.687^{+0.241}_{-0.221}$ & $7.083^{+0.232}_{-0.226}$ & $6.712^{+0.205}_{-0.215}$ \\
            $\rho_{s}$ [kg~m$^{-3}$] & $1385.1^{+118.4}_{-103.2}$ & $679.0^{+68.8}_{-62.9}$ & $543.0^{+51.2}_{-50.4}$ \\
            e & 0 (fixed) & 0 (fixed) & 0 (fixed) \\
            $\omega$ [deg] & 90 (fixed) & 90 (fixed) & 90 (fixed) \\
            K [m~s$^{-1}$] & $122.6^{+15.4}_{-15.3}$ & - & $40.1^{+8.5}_{-8.7}$ \\
            \midrule
			\headrow \multicolumn{4}{|c|}{\textbf{Derived Properties}} \\
            \midrule
            $R_{p}$ [$R_{J}$] & $1.21 \pm 0.03$ & $1.15 \pm 0.05$ & $1.24 \pm 0.04$ \\
            $M_{p}$ [$M_{J}$] & $0.85 \pm 0.12$ & - & $0.30 \pm 0.07$ \\
            $\rho_{p}$ [g~cm$^{-3}$] & $0.64 \pm 0.10$ & - & $0.21 \pm 0.05$ \\
            i [deg] & $87.3 \pm 0.5$ & $83.3 \pm 0.4$ & $83.0 \pm 0.4$ \\
            a [au] & $0.039 \pm 0.001$ & $0.042 \pm 0.002$ & $0.043 \pm 0.002$ \\
            $T_{eq}$ ($A_{B}$ = 0) [K] & $1290 \pm 28$ & $1469 \pm 32$ & $1561 \pm 35$ \\
            F [W~m$^{-2}$] & (6.3 $\pm$ 0.7)$\times 10^5$ & (1.1 $\pm$ 0.1) $\times 10^6$ & (1.4 $\pm$ 0.1) $\times 10^6$ \\
            \midrule 
			\headrow \multicolumn{4}{|c|}{\textbf{Instrumental Parameters}} \\
            \midrule 
            $M_{TESS600}$ & -0.001 $\pm$ 0.020 & 0.002 $\pm$ 0.006 &  -0.0001 $\pm$ 0.0001 \\
            $M_{TESS200}$ & 0.012 $\pm$ 0.031  & 0.008 $\pm$ 0.009 & -0.0004 $\pm$ 0.0003 \\
            $M_{H50r}$ & -0.057 $\pm$ 0.064 & -0.009 $\pm$ 0.031 & -0.005 $\pm$ 0.016\\
            $M_{H50V}$ & -0.001 $\pm$ 0.087 & -0.006 $\pm$ 0.012 & -0.008 $\pm$ 0.050 \\
            $M_{H50g}$ & -0.012 $\pm$ 0.002 & - & - \\
            $M_{H50/CTIOi}$ & - & -0.003 $\pm$ 0.013 & -0.0003 $\pm$ 0.0100 \\
            $J_{TESS600}$ [ppm] & $2.7^{+32.2}_{-2.4}$ & $8.4^{+48.3}_{-7.7}$ & $5.2^{+48.5}_{-4.7}$ \\
            $J_{TESS200}$ [ppm] & $3.6^{+44.1}_{-3.3}$ & $2.1^{+22.8}_{-1.8}$ & $31.0^{+89.6}_{-27.7}$ \\
            $J_{H50r}$ [ppm] & $11.0^{+158.6}_{-10.5}$ & $7.3^{+140.6}_{-6.8}$ & $16.8^{+182.5}_{-15.8}$ \\
            $J_{H50V}$ [ppm] & $10.2^{+166.5}_{-9.7}$ & $28.6^{+254.7}_{-27.3}$ & $23.7^{+245.6}_{-22.5}$ \\
            $J_{H50g}$ [ppm] & $14.2^{+523.6}_{-13.7}$ & - & - \\
            $J_{H50/CTIOi}$ [ppm] & - & $7.69^{+152.5}_{-7.27}$ & $26.15^{+236.15}_{-24.83}$\\
            $D_{H50r}$ & 1.0 (fixed) & $0.95 \pm 0.03$ & 1.0 (fixed) \\
            $D_{H50V}$ & 1.0 (fixed) & $0.96 \pm 0.03$ & 1.0 (fixed) \\
            $D_{H50g}$ & 1.0 (fixed) & - & - \\
            $D_{H50/CTIOi}$ & - & $0.95 \pm 0.03$ & 1.0 (fixed) \\
            $q_{1, TESS}$ & 0.29 $\pm$ 0.03 & 0.30 $\pm$ 0.03 & 0.28 $\pm$ 0.03\\
            $q_{2, TESS}$ & 0.38 $\pm$ 0.04 & 0.39 $\pm$ 0.04 & 0.37 $\pm$ 0.04 \\
            $q_{1, H50r}$ & 0.42 $\pm$ 0.04 & 0.41 $\pm$ 0.04 & 0.38 $\pm$ 0.04\\
            $q_{2, H50r}$ & 0.40 $\pm$ 0.04 & 0.41 $\pm$ 0.04 & 0.40 $\pm$ 0.04\\
            $q_{1, H50V}$ & 0.47 $\pm$ 0.05 & 0.46 $\pm$ 0.05 & 0.47 $\pm$ 0.05\\
            $q_{2, H50V}$ & 0.44 $\pm$ 0.04 & 0.42 $\pm$ 0.04 & 0.41 $\pm$ 0.04\\
            $q_{1, H50g}$ & 0.60 $\pm$ 0.06 & - & - \\
            $q_{2, H50g}$ & 0.48 $\pm$ 0.05 & - & - \\
            $q_{1, H50/CTIOi}$ & - & 0.30 $\pm$ 0.03 & 0.29 $\pm$ 0.02 \\
            $q_{2, H50/CTIOi}$ & - & 0.39 $\pm$ 0.04 & 0.36 $\pm$ 0.03 \\
            $\mu_{CHIRON/FEROS}$ [m~s$^{-1}$] & $49555.91^{+11.31}_{-12.07}$ & - & $32976.16^{+8.32}_{-8.97}$ \\
            $J_{CHIRON/FEROS}$ [m~s$^{-1}$] & $0.20^{+3.94}_{-0.19}$ & - & $46.50^{+7.45}_{-6.15}$ \\
            $\mu_{HARPS}$ [m~s$^{-1}$] & - & - & $32978.44^{+9.76}_{-10.42}$ \\
            $J_{HARPS}$ [m~s$^{-1}$] & - & - & $0.28^{+7.88}_{-0.27}$ \\
			\bottomrule 
		\end{tabular}
		\caption{The modelling results from Juliet for the three TASSIE II systems.}
		\label{table:planetresults}
	\end{center}
\end{table*}

\section{Discussion}
\subsection{TASSIE II TOIs and the Short-Period Planet Population}
\begin{figure*}[h!]
    \begin{subfigure}[]
    \centering
    \includegraphics[height = 0.45\linewidth, width=0.75\linewidth]{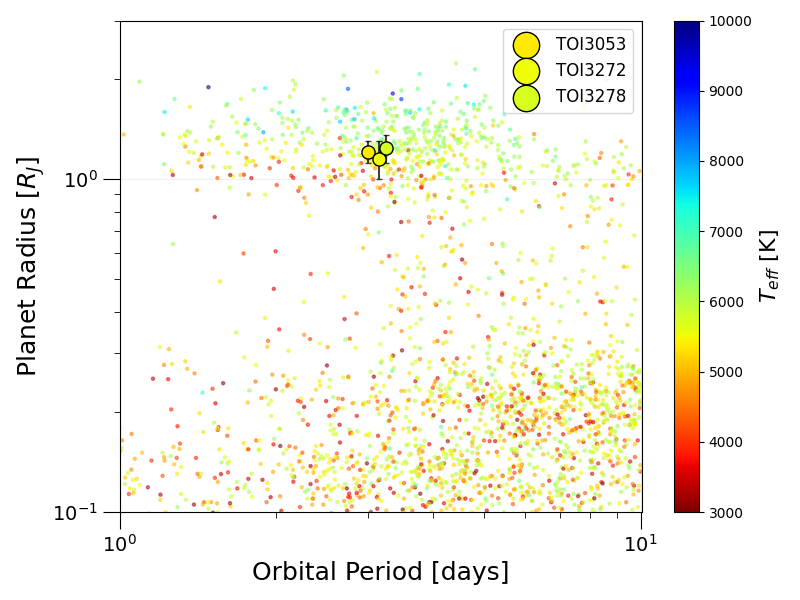}
    \end{subfigure}
    ~
    \begin{subfigure}[]
    \centering
    \includegraphics[height = 0.45\linewidth, width=0.7\linewidth]{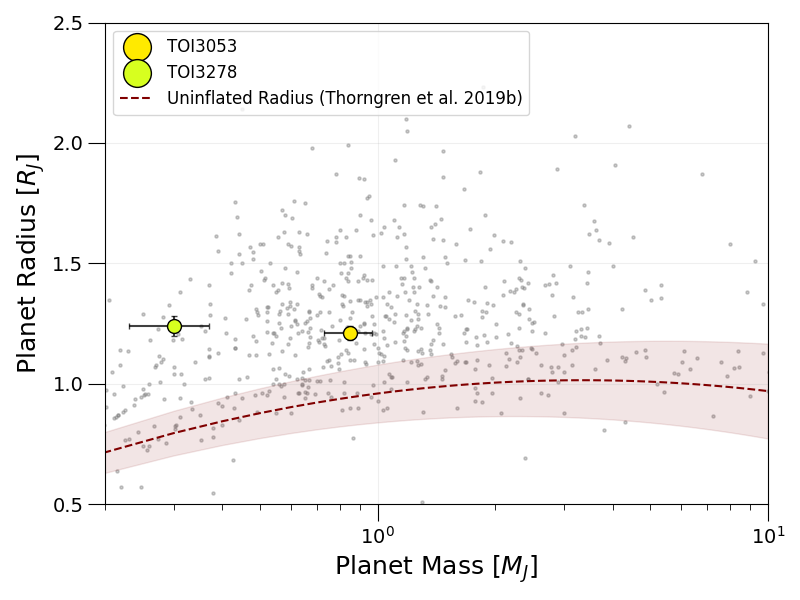}
    \end{subfigure}
    ~
    \caption{The mass-radius-period parameter spaces for short-period exoplanets. In \textbf{(a)}, the radii and orbital periods of known planets are shown with systems coloured by the stellar effective temperature. The TASSIE II TOIs are highlighted as points with $3\sigma$ error bars for their radii. In \textbf{(b)}, the radii and masses are shown for known HJs. TOI-3053b and TOI-3278b / HATS-78b are shown with error bars representing the $1\sigma$ errors on mass. The dashed line and shaded regions show the predicted uninflated radius as a function of mass using the relation from \citet{thorngren2019b}. Data from the NASA Exoplanet Archive.}
    \label{fig:MvR}
\end{figure*}

We now consider the TASSIE II systems in context of the broader short-period planet population\footnote{Data from: \url{https://exoplanetarchive.ipac.caltech.edu}}. As shown in Figure \ref{fig:MvR} (a), TOI-3272.01 appears consistent with being a HJ based on the position in the radius-period parameter space at the well-known `three day pile-up', sitting with the other two planets studied in this work. For TOI-3053b and TOI-3278b / HATS-78b, we can gain further insights by comparing the measured radii with expectations from empirical mass-radius relations. We can see from Figure \ref{fig:MvR}b that both TOI-3053b and TOI-3278b / HATS-78b appear to be inflated, sitting above the uninflated radius line from \citet{thorngren2019b}. This relation was derived from a sample of `warm/cool' giants (with orbital periods greater than 10 days), including those from the Solar System. It has long been noted that most HJs are inflated, with a correlation between the radius anomaly and incident flux \citep[e.g.,][]{weiss2013}. A universal explanation for radius inflation remains elusive, but mechanisms such as ohmic heating or thermal tides have been shown to each match certain features of the observed HJ population \citep[see][]{inflation}. A larger sample of hot giant planets, especially around young and post-main sequence stars, will be required to determine which mechanism is the predominant avenue of inflation. Interestingly, the inferred age of TOI-3278 implies that is is close to the end of its main-sequence lifetime ($\approx$ 9 - 10 Gyr). The proximity of the star to the main-sequence turn-off in the Hertzsprung-Russell diagram from SED + MIST isochrone fitting is consistent with this scenario (as shown in Figure \ref{fig:HR_Diagram_TOI3278}). This would imply that TOI-3278b / HATS-78b may be undergoing a period of re-inflation, as the expansion of its host star increases the irradiation of the planet. This makes the system an intriguing target for further study.  

\subsection{Future Observations and Atmospheric Characterisation}
Whilst we were able to confirm TOI-3053b and TOI-3278b / HATS-78b as planetary in nature, TOI-3272.01 remains as a candidate despite passing various false positive tests. The RV signal from TOI-3272 will likely be more challenging to characterise, due to the presence of spots/activity and the host star being relatively faint. Confirming that TOI-3272 is indeed a young star will be an important first step to decide if RV observations should be attempted, with a single high-resolution spectrum targeting typical youth indicators (e.g., Ca H \& K line emission or Li non-depletion) being recommended. If this target is indeed a giant planet orbiting a young field star, this could be an important addition to our understanding of giant planet evolution through the first billion years of a system's life. High angular resolution imaging of TOI-3272 and TOI-3278 would also be beneficial to firmly rule out any stellar companions below the \gaia detection limit that may contaminate photometry and influence the radius and age estimates. 
\newline\newline
We also assessed each target's suitability for atmospheric measurements with \textit{JWST} or \textit{Ariel}. We found that the targets are generally too faint to perform transmission spectroscopy with either instrument, with some caveats. For TOI-3053b and TOI-3278b / HATS-78b, we calculate the Transmission Spectroscopy Metric (TSM) following \citet{tsm_kempton} and find $TSM_{3053} = 70 \pm 12$  and $TSM_{3278} = 71 \pm 18$ using our previously derived parameters. For TOI-3272.01 which does not have a mass measurement, we assume a fiducial mass of 1 $M_{J}$ and estimate a TSM of 24. These all fall below the recommended threshold of TSM > 90 for giant planets \citep{tsm_kempton}. However, if the mass of either is closer to their lower confidence interval bounds, then the planets will pass this threshold and become possible targets for transmission spectroscopy with \textit{JWST}. Regardless, the low density and possible re-inflation of TOI-3278b / HATS-78b may still justify investigation with \textit{JWST} phase curves to map the temperature profile of the planet and look for signatures of atmospheric escape \citep[e.g., through infrared metastable He absorption, see][]{wasp107b}. Furthermore, the upcoming release of \gaia DR4 epoch  astrometry presents an exciting opportunity to search for long-period outer companions to these systems, as expected from a HEM migration channel \citep[e.g.,][]{DR4Yield}.

\section{Conclusion}
In this paper, we have studied three close-in giant transiting exoplanet candidates alerted by \textit{TESS} that are orbiting Sun-like stars. Using ground-based follow-up light curves from the UTGO H50 telescope, along with auxiliary astrometric, imaging and RV data, we were able to confirm TOI-3053b and TOI-3278b / HATS-78b as hot giant planets. TOI-3053b is a typical hot Jupiter, with a radius of $R_{3278b}$ = 1.21 $\pm$ 0.03 $R_{J}$ and mass of $M_{3278b}$ = 0.85 $\pm$ 0.12 $M_{J}$. TOI-3278b / HATS-78b was found to be a highly inflated Saturn-mass planet ($R_{3278b}$ = 1.24 $\pm$ 0.04 $R_{J}$, $M_{3278b}$ = 0.30 $\pm$ 0.07 $M_{J}$) orbiting a potentially old star (T$_{3278}$ = 8.9$^{+2.9}_{-2.3}$ Gyr) that could be close to the end of the main-sequence lifetime. We also investigated TOI-3272.01 which, despite remaining unconfirmed, is consistent with being a hot Jupiter (with $R_{3272.01} = 1.15 \pm 0.05 \, R_{J}$). TOI-3272 is an exciting target for further study as a potentially young system (T$_{3272}$ = $1.1 \pm 0.2$ Gyr), which may provide an important constraint for studies of the radius inflation of HJs. Whilst these planets are not ideal targets for atmospheric measurements with \textit{JWST} or \textit{Ariel}, future studies using \gaia DR4 epoch astrometry and \textit{JWST} phase curves remain a viable possibility. Furthermore, these systems add to our sample of companions orbiting relatively faint stars (V > 13 mag) that will be common targets for upcoming space-based transit missions, such as the Nancy Grace Roman Galactic Bulge Time Domain Survey \citep{roman}. Although these faint and distant host stars present challenges to confirmation efforts \citep{roman_transits}, they will be vital in understanding the influence of galactic environment on exoplanet formation and evolution.  

\begin{acknowledgement}
The authors wish to thank both Andrew Vanderburg for his valuable suggestions for the false positive vetting of TOI-3053.02 and Kim McLeod for providing one of the LCO datasets used in this work. We also thank the anonymous reviewer for their comments that helped to improve the manuscript.

Based on observations obtained at the Greenhill Observatory, which is owned and operated by the University of Tasmania. The observatory was made possible by generous donations from Caisey Harlingten and the UTAS Foundation. We acknowledge Palawa, as the original owners of Lutruwita/Tasmania on which the observatory stands.

This work uses data acquired at the Siding Spring Observatory with the Australian National University 2.3m Telescope. We acknowledge the traditional custodians of the land on which the telescope stands, the Gamilaraay people, and pay our respects to elders past and present. 

This paper includes data collected by the \textit{TESS} mission, which are
publicly available from the Mikulski Archive for Space Telescopes
(MAST). Funding for the \textit{TESS} mission is provided by NASA’s
Science Mission directorate. We acknowledge the use of public
\textit{TESS} Alert data from pipelines at the \textit{TESS} Science Office and at
the \textit{TESS} SPOC. Resources supporting this work were provided
by the NASA High-End Computing (HEC) Programme through
the NASA Advanced Supercomputing (NAS) Division at Ames
Research Centre for the production of the SPOC data products. 

This work has made use of data from the European Space Agency
(ESA) mission \gaia (\url{https://www.cosmos.esa.int/gaia}), processed
by the Gaia Data Processing and Analysis Consortium (DPAC,  \url{https:
//www.cosmos.esa.int/web/gaia/dpac/consortium}). Funding for the
DPAC has been provided by national institutions, in particular
the institutions participating in the \gaia Multilateral Agreement. 

This research has made use of the SIMBAD data base and VizieR
catalogue access tool, operated at CDS, Strasbourg, France. 

This research has made use of NASA’s Astrophysics Data System. This research has made use of the NASA Exoplanet Archive, which is
operated by the California Institute of Technology, under contract
with the National Aeronautics and Space Administration under the
Exoplanet Exploration Programme. 

\end{acknowledgement}

\paragraph{Funding Statement}
T.P. is supported by an Australian Government Research Training Program (RTP) Scholarship. E.T. has been supported by the Australian Government through the Australian Research Council Discovery Project Grant 240101842. J.S. was supported in part by funding provided by the National Aeronautics and Space Administration (NASA), under award number 80NSSC20M0124, Michigan Space Grant Consortium (MSGC). K.P. acknowlegdes support from NSF grant 2311527, and NASA grant 80NSSC25K0379. R.B. acknowledges support from Fondecyt Project 1241963. A.J. acknowledges support from Fondecyt project 1251439. Development of the HATSouth project was funded by NSF MRI grant NSF/AST-0723074, operations were supported by NASA grants NNX09AB29G, NNX12AH91H, and \newline NNX17AB61G, and follow-up observations were partially supported from grant NSF/AST-1108686. 

\paragraph{Competing Interests}
The authors are not aware of any competing interests.

\paragraph{Data Availability Statement}
All light curves and spectra obtained for the TASSIE program will be uploaded to ExoFOP (\url{https://exofop.ipac.caltech.edu/tess/}) after publishing. FEROS and HARPS spectra are available through the ESO Science Archive. The analysis notebooks for this work can be viewed here: \url{https://github.com/tjplunkett/TASSIE_II}

\paragraph{Author Contributions}
Conceptualization: T.P. Methodology: T.P; E.T; J.R.; A.C.; JP.B. Observations/Data Curation: T.P; A.C; J.S; B.E; + remaining authors. Data visualisation: T.P; C.Z. Writing original draft: T.P. All authors approved the final submitted draft.

\FloatBarrier

\bibliography{ref.bib}

\appendix
\setcounter{figure}{0}
\renewcommand{\thefigure}{A\arabic{figure}}
\setcounter{table}{0}
\renewcommand{\thetable}{A\arabic{table}}

\section{Auxiliary Figures and Tables}
\begin{figure}
    \centering
    \includegraphics[width=1\linewidth, height = 0.68\linewidth]{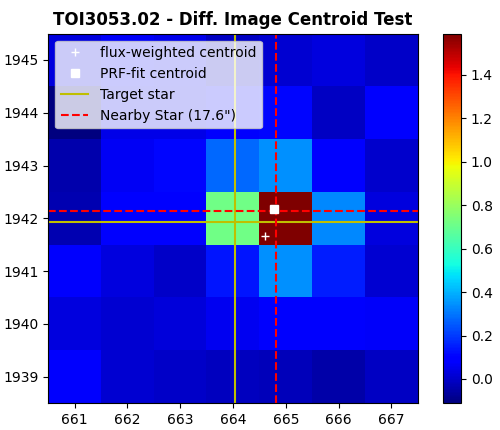}
    \caption{The difference image centroid test for the 0.416 day signal in the TOI-3053 light curves. The positions of TOI-3053 and the nearby false positive EB are highlighted with gold and red crosses, respectively. The colour bar indicates the decrease in flux measured in-eclipse (in parts per thousand).}
    \label{fig:centroid}
\end{figure}

\begin{figure}[h!]
    \centering
    \includegraphics[width=1\linewidth, height=0.64\linewidth]{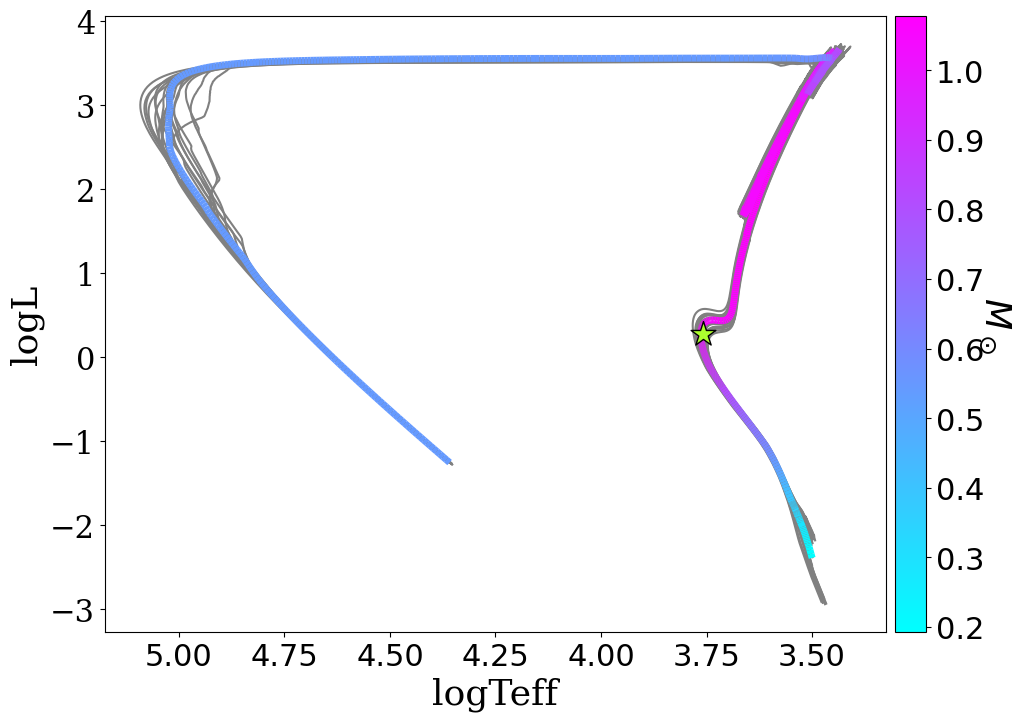}
    \caption{The Hertzsprung-Russell diagram of TOI-3278 / HATS-78 from the ARIADNE SED + MIST isochrone fitting. The thick coloured line is the interpolated MIST isochrone corresponding to the best-fit stellar parameters. The colourbar traces the stellar mass for each section of the evolutionary track. The grey lines are 20 random samples drawn from the parameter posterior distribution, showing the uncertainty on the fit. The star icon shows the inferred logarithm of the luminosity and temperature for TOI-3278 / HATS-78, with the position close to the main-sequence turn-off.}
    \label{fig:HR_Diagram_TOI3278}
\end{figure}

\begin{table*}[h!]
	\begin{center}
		\begin{tabular}{|l|p{3.2cm}|p{1.8cm}|p{3cm}|p{2.75cm}|p{0.85cm}|p{0.7cm}|}
			\toprule
			\headrow \textbf{ID} & \textbf{Instrument} & \textbf{Camera} & \textbf{Pixel Scale [arcsec~pix$^{-1}$]} & \textbf{Dates [UTC]} & 
            \textbf{Filter} & \textbf{N$_{Images}$} \\
			\midrule 
            \textbf{TOI-3053} & Brierfield 0.36 m \newline Evans 0.51 m & Moravian 16803 \newline STT 1603-3 & 1.47 (2$\times$2 bin) \newline 1.08 (2$\times$2 bin) & 2021/09/08 \newline 2022/02/14 & R \newline B & 49 \newline 93 \\
            \hline 
			\textbf{TOI-3278 / HATS-78} & HATSouth 180 mm Network \newline LCO 1 m (SSO) \newline LCO 1 m (SSO) \newline LCO 1 m (CTIO) & Apogee U16m \newline  SBIG \newline Sinistro \newline Sinistro & 3.7 \newline 0.23 \newline 0.39 \newline 0.39
            & 2011/04/26 - 2012/11/20 2015/07/06 \newline 2015/07/26 \newline 2021/06/13
            & SDSS $r^{\prime}$ \newline SDSS $i^{\prime}$
            \newline SDSS $i^{\prime}$ \newline SDSS $i^{\prime}$ & 9062 \newline 71 \newline 74 \newline 173 \\
			\bottomrule 
		\end{tabular}
		\caption{Additional photometry logs for TASSIE II TOIs.}
		\label{table:tassiephotlog2}
	\end{center}
\end{table*}

\begin{figure}[h!]
     \begin{subfigure}[]
	   \centering
       \includegraphics[width=0.95\linewidth]{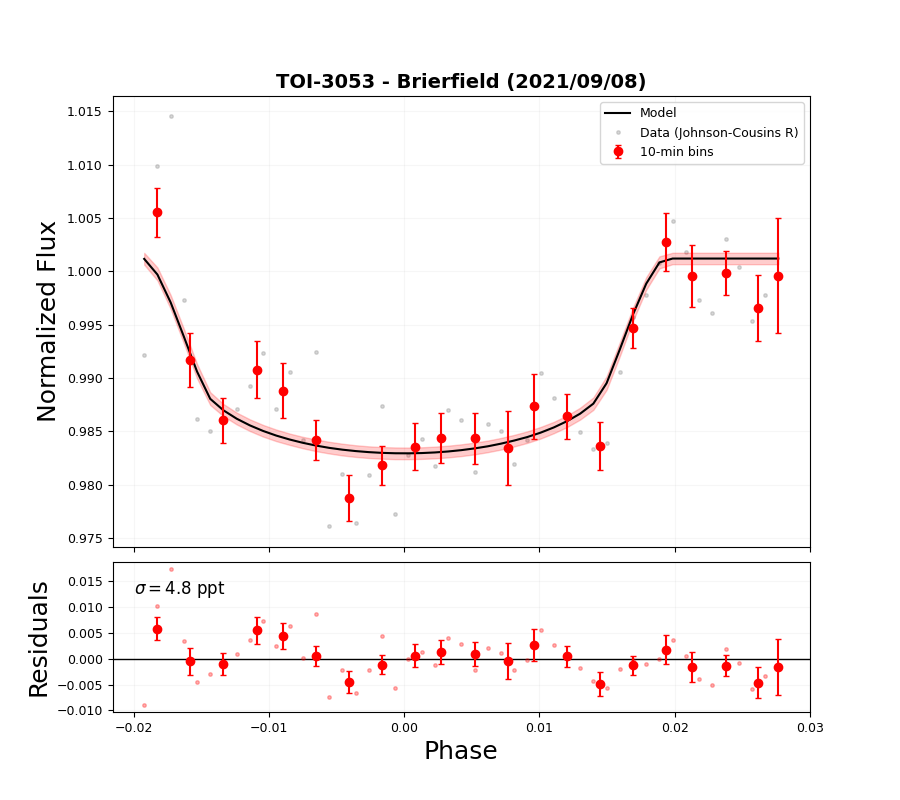}
    \end{subfigure}
    ~
    \begin{subfigure}[]
	   \centering
        \includegraphics[width=0.95\linewidth]{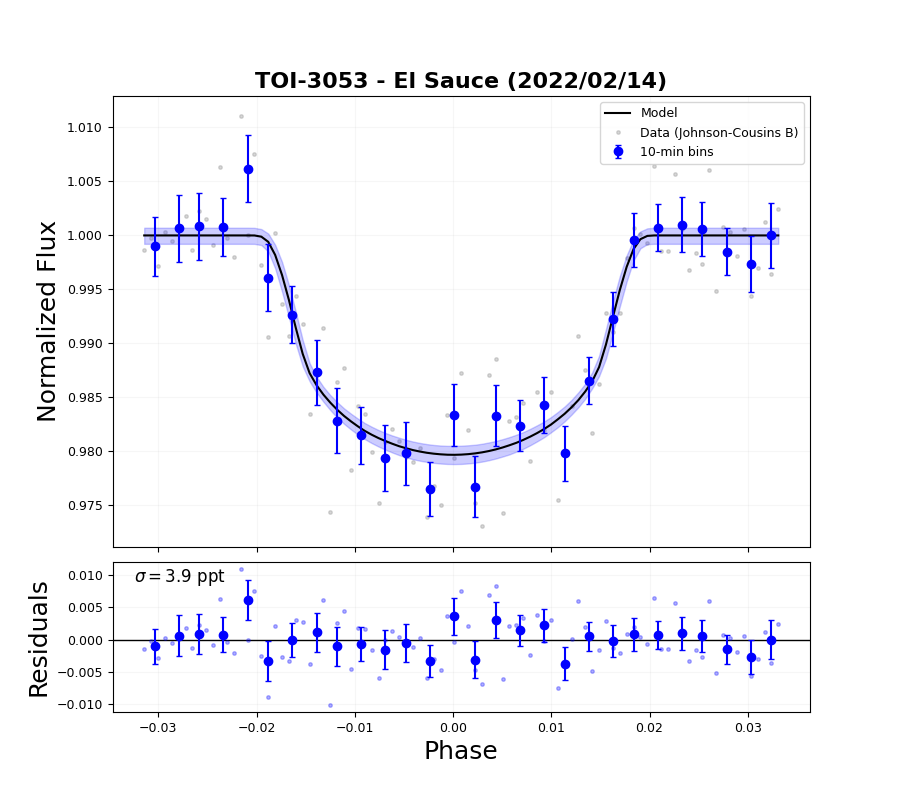}
    \end{subfigure}
    \caption{Follow-up light curves for TOI-3053b from the Brierfield (\textbf{a}) and El Sauce (\textbf{b}) observatories. The coloured points with error bars show the individual photometric points (in grey) binned to 10-minutes for clarity. The best-fit transit model from Section 4.2 is overlaid in black. The residuals and photometric scatter are shown in the bottom panels of each plot.}
    \label{fig:TOI3053SG1}
\end{figure}

\begin{table}[h!]
	\begin{center}
		\begin{tabular}{|l|l|l|}
        \toprule
        \textbf{Instrument} & \textbf{BJD-TDB} [days] & \textbf{RV} [m~s$^{-1}$] \\
        \midrule
        \headrow \multicolumn{3}{|c|}{\textbf{TOI-3053}} \\
        \midrule 
        CHIRON & 2461136.545220 & 49422.0$^{+51.0}_{-51.0}$\\
        CHIRON & 2461137.591260 & 49601.0$^{+51.0}_{-51.0}$ \\
        CHIRON & 2461138.630000 &   49616.0$^{+47.0}_{-47.0}$ \\
        CHIRON & 2461139.532130 & 49429.0$^{+33.0}_{-33.0}$ \\
        CHIRON & 2461140.589780 & 49659.0$^{+46.0}_{-46.0}$ \\
        CHIRON & 2461142.574960 & 49454.0$^{+36.0}_{-36.0}$ \\
        CHIRON & 2461144.528690 & 49559.0$^{+34.0}_{-34.0}$ \\
        CHIRON & 2461145.581920 & 49426.0$^{+39.0}_{-39.0}$ \\
        CHIRON & 2461146.559430 & 49648.0$^{+30.0}_{-30.0}$ \\
        CHIRON & 2461147.625680 & 49594.0$^{+38.0}_{-38.0}$ \\
        \midrule
        \headrow \multicolumn{3}{|c|}{\textbf{TOI-3278 / HATS-78}} \\
        \midrule
        FEROS & 2457183.618548 & 33003.0$^{+18.0}_{-18.0}$ \\
        FEROS & 2457184.820261 & 32926.0$^{+18.0}_{-18.0}$ \\
        FEROS & 2457186.693816 & 33045.0$^{+21.0}_{-21.0}$ \\
        FEROS & 2457189.908373 & 33031.0$^{+16.0}_{-16.0}$ \\
        FEROS & 2457191.820070 & 32971.0$^{+19.0}_{-19.0}$ \\
        FEROS & 2457194.780835 & 32984.0$^{+21.0}_{-21.0}$ \\
        FEROS & 2457225.523469 & 33029.0$^{+15.0}_{-15.0}$ \\
        FEROS & 2457228.817121 & 32924.0$^{+23.0}_{-23.0}$ \\
        FEROS & 2457234.847000 & 33104.0$^{+33.0}_{-33.0}$ \\
        FEROS & 2457235.546146 & 33029.0$^{+16.0}_{-16.0}$ \\
        FEROS & 2457238.564891 & 32992.0$^{+21.0}_{-21.0}$ \\
        FEROS & 2457557.775233 & 32936.0$^{+12.0}_{-12.0}$ \\
        FEROS & 2457558.572383 & 32890.0$^{+14.0}_{-14.0}$\\
        FEROS & 2457575.590787 & 33060.0$^{+16.0}_{-16.0}$ \\
        HARPS & 2457637.525480 & 32961.3$^{+34.6}_{-34.6}$ \\
        HARPS & 2457639.548477 & 32932.2$^{+27.1}_{-27.1}$ \\
        HARPS & 2457640.500189 & 33006.4$^{+13.5}_{-13.5}$ \\
        FEROS & 2457835.818547 & 33035.7$^{+17.3}_{-17.3}$\\
        FEROS & 2457837.881704 & 33037.0$^{+15.7}_{-15.7}$\\
        FEROS & 2457844.839364 & 32898.6$^{+15.6}_{-15.6}$\\
        FEROS & 2457903.831382 & 32925.9$^{+21.3}_{-21.3}$ \\
        FEROS & 2457904.740707 & 32979.7$^{+15.6}_{-15.6}$ \\
        FEROS & 2457906.816860 & 33039.2$^{+13.7}_{-13.7}$\\
        FEROS & 2457908.717873 & 32899.9$^{+15.5}_{-15.5}$\\
        FEROS & 2457909.677443 & 32981.6$^{+15.8}_{-15.8}$ \\
        FEROS & 2457910.795380 & 33001.2$^{+16.9}_{-16.9}$ \\
        FEROS & 2457911.819862 & 32951.6$^{+19.3}_{-19.3}$\\
        \bottomrule
        \end{tabular}
		\caption{Radial velocities for TOI-3053 and TOI-3278 / HATS-78.}
		\label{table:rvlogs}
	\end{center}
\end{table}

\FloatBarrier

\begin{figure*}
    \centering
    \includegraphics[width=0.95\linewidth, height = 0.55\linewidth]{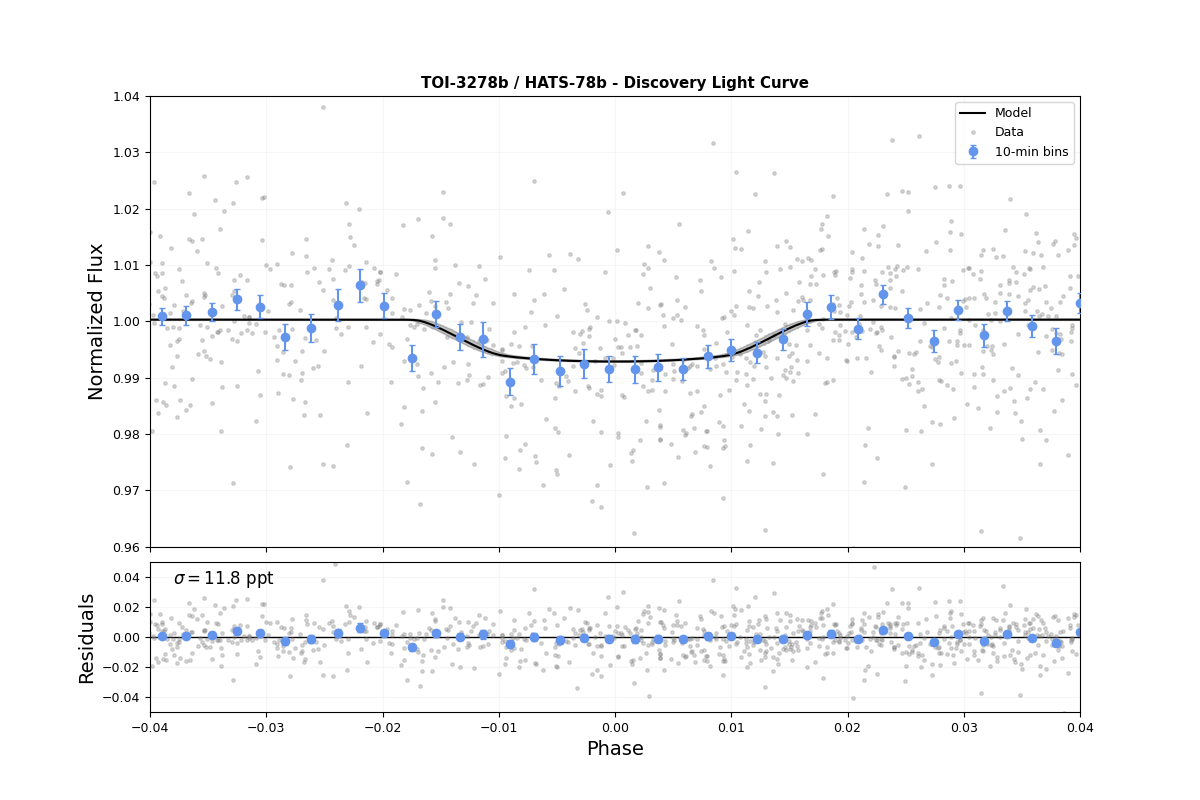}
    \caption{The discovery light curve for TOI-3278b / HATS-78b from the HATSouth telescope network. The photometric measurements are shown as the grey points, with best-fit model derived from the global fit in Section 4.2 plotted in black. The blue points show the data binned to 10-minutes for clarity. The residuals are plotted in the lower panel, showing the photometric scatter of 11.8 ppt.}
    \label{fig:HATSLCTOI-3278b}
\end{figure*}

\begin{figure*}
     \begin{subfigure}[]
	   \centering
       \includegraphics[width=0.49\linewidth]{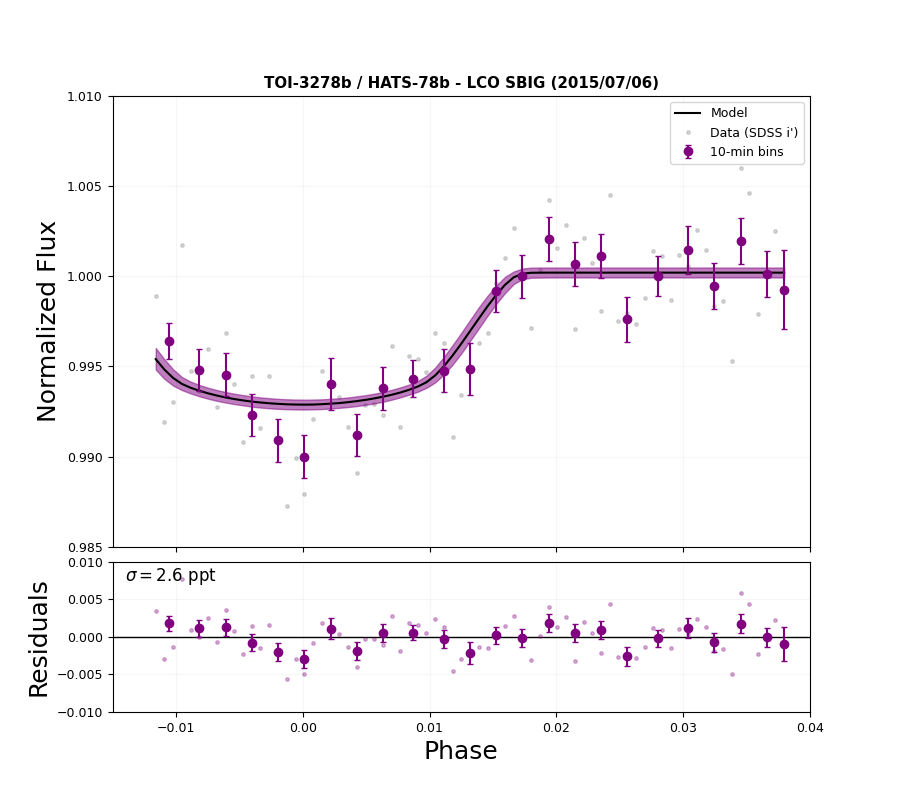}
    \end{subfigure}
    ~
    \begin{subfigure}[]
	   \centering
        \includegraphics[width=0.49\linewidth]{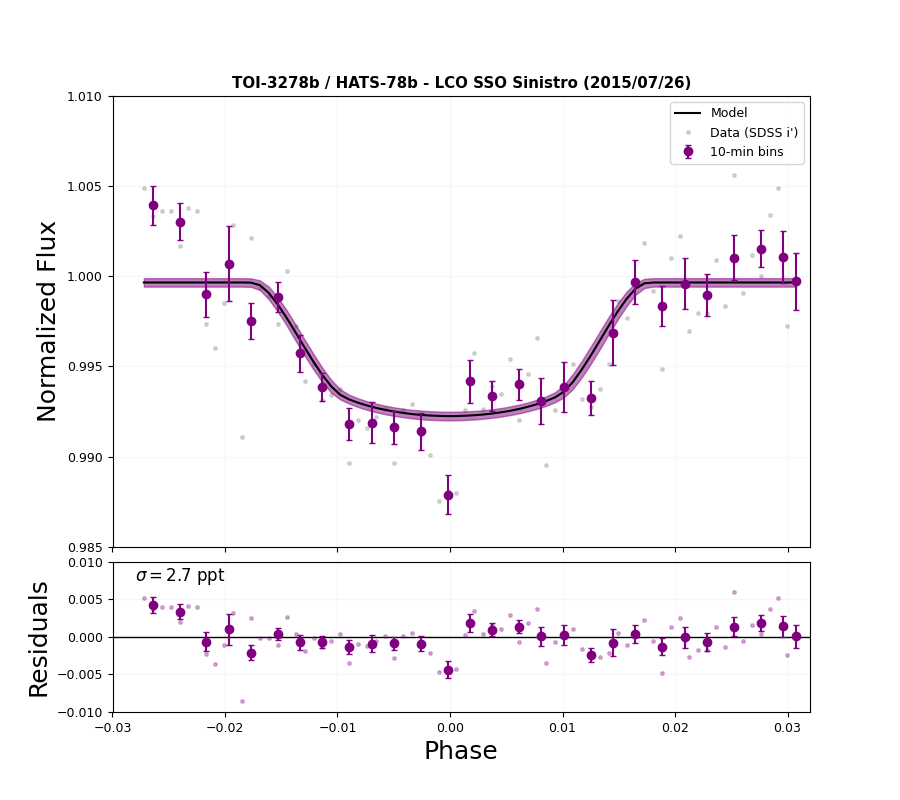}
    \end{subfigure}
    \caption{Follow-up light curves for TOI-3278b / HATS-78b obtained with the LCO 1m telescopes at SSO on 2015/07/06 \textbf{(a)} and 2015/07/26 \textbf{(b)}. The SDSS $i^{\prime}$ photometry is shown as the grey points, with the 10-minute binned data in purple and best-fit model from global fitting in Section 4.2 overlaid in black. The lower panels shows the residuals and photometric scatter. Note, these light curves were not used in the global fitting stage.}
    \label{fig:LCOFOLLOWUPTOI-3278b}
\end{figure*}

\FloatBarrier 

\begin{table*}
    \begin{tabular}{|l|l|l|l|l|}
    \toprule
    \textbf{Parameter} & \textbf{Description} & \textbf{TOI-3053} & \textbf{TOI-3272} & \textbf{TOI-3278} \\
    \midrule
    $P_{p1}$ [days] & Orbital period &  $\mathcal{N}$(2.99, 0.05) & $\mathcal{N}$(3.15, 0.05) & $\mathcal{N}$(3.25, 0.1) \\
    $t_{0,p1}$ [BJTD - 2457000] & Transit midtime & $\mathcal{N}$(2358.51, 0.05) & $\mathcal{N}$(2387.06, 0.05) & $\mathcal{N}$(1680.81, 0.05) \\
    $r1_{p1}$ & Parametrisation for radius ratio and impact factor & $\mathcal{U}$(0.0, 1.0) & $\mathcal{U}$(0.0, 1.0) & $\mathcal{U}$(0.0, 1.0) \\
    $r2_{p1}$ & "" & $\mathcal{U}$(0.0, 1.0) & $\mathcal{U}$(0.0, 1.0) & $\mathcal{U}$(0.0, 1.0)\\
    K [m~s$^{-1}$] & Radial velocity amplitude & - & - & $\mathcal{U}$(0, 1000) \\
    $\rho_s$ [kg~m$^{-3}$] & Stellar density & $\mathcal{N}$(1481.0, 153.7) & $\mathcal{N}$(673.6, 72.8) & $\mathcal{N}$(545.3, 63.2) \\
    $q1_{TESS}$ & Limb darkening parameter for TESS & $\mathcal{N}$(0.30, 0.03) & $\mathcal{N}$(0.30, 0.03) & $\mathcal{N}$(0.28, 0.03) \\
    $q2_{TESS}$ & "" & $\mathcal{N}$(0.38, 0.04) & $\mathcal{N}$(0.38, 0.04) & $\mathcal{N}$(0.37, 0.04) \\
    $q1_{H50r}$ & Limb darkening parameter for SDSS $r^{\prime}$ & $\mathcal{N}$(0.42, 0.04) & $\mathcal{N}$(0.41, 0.04) & $\mathcal{N}$(0.38, 0.04) \\
    $q2_{H50r}$ & "" & $\mathcal{N}$(0.40, 0.04) & $\mathcal{N}$(0.40, 0.04) & $\mathcal{N}$(0.40, 0.04) \\
    $q1_{H50g}$ & Limb darkening parameter for SDSS $g^{\prime}$ & $\mathcal{N}$(0.58, 0.06) & - & - \\
    $q2_{H50g}$ & "" & $\mathcal{N}$(0.47, 0.05) & - & - \\
    $q1_{H50V}$ & Limb darkening parameter for Bessell V & $\mathcal{N}$(0.48, 0.05) & $\mathcal{N}$(0.47, 0.05) & $\mathcal{N}$(0.45, 0.05) \\
    $q2_{H50V}$ & "" & $\mathcal{N}$(0.43, 0.04) & $\mathcal{N}$(0.42, 0.04) & $\mathcal{N}$(0.41, 0.04) \\
    $q1_{H50/CTIOi}$ & Limb darkening parameter for SDSS $i^{\prime}$ & - & $\mathcal{N}$(0.30, 0.03) &  $\mathcal{N}$(0.28, 0.03) \\
    $q2_{H50/CTIOi}$ & "" & - & $\mathcal{N}$(0.38, 0.04) & $\mathcal{N}$(0.37, 0.04) \\
    $M_{TESS600}$ & Flux offset for TESS (600 s) & $\mathcal{N}$(0.0, 0.1) & $\mathcal{N}$(0.0, 0.1) & $\mathcal{N}$(0.0, 0.1) \\
    $J_{TESS600}$ [ppm] & Photometric jitter for TESS (600 s) & $log\mathcal{U}$(0.1, 1000.0) & $log\mathcal{U}$(0.1, 1000.0) & $log\mathcal{U}$(0.1, 1000.0) \\
    $M_{TESS200}$ & Flux offset for TESS (200 s) & $\mathcal{N}$(0.0, 0.1) & $\mathcal{N}$(0.0, 0.1) & $\mathcal{N}$(0.0, 0.1) \\
    $J_{TESS200}$ [ppm] & Photometric jitter for TESS (200 s) & $log\mathcal{U}$(0.1, 1000.0) & $log\mathcal{U}$(0.1, 1000.0) & $log\mathcal{U}$(0.1, 1000.0) \\
    $D_{H50r}$ & Dilution factor for SDSS $r^{\prime}$ & 1.0 & $\mathcal{U}$(0.9, 1.0) & 1.0 \\
    $M_{H50r}$ & Flux offset for SDSS $r^{\prime}$ & $\mathcal{N}$(0.0, 0.1) & $\mathcal{N}$(0.0, 0.1)  & $\mathcal{N}$(0.0, 0.1)  \\
    $J_{H50r}$ [ppm] & Photometric jitter for SDSS $r^{\prime}$ &$log\mathcal{U}$(0.1, 1000.0) & $log\mathcal{U}$(0.1, 1000.0) & $log\mathcal{U}$(0.1, 1000.0)\\
    $D_{H50g}$ & Dilution factor for SDSS $g^{\prime}$ & 1.0 & - & - \\
    $M_{H50g}$ & Flux offset for SDSS $g^{\prime}$ & $\mathcal{N}$(0.0, 0.1) & - & - \\
    $J_{H50g}$ [ppm] & Photometric jitter for SDSS $r^{\prime}$ & $log\mathcal{U}$(0.1, 1000.0) & - & - \\
    $D_{H50V}$ & Dilution factor for Bessell V & 1.0 & $\mathcal{U}$(0.9, 1.0) & 1.0 \\
    $M_{H50V}$ & Flux offset for Bessell V & $\mathcal{N}$(0.0, 0.1) & $\mathcal{N}$(0.0, 0.1)  & $\mathcal{N}$(0.0, 0.1)  \\
    $J_{H50V}$ [ppm] & Photometric jitter for Bessell V & $log\mathcal{U}$(0.1, 1000.0) & $log\mathcal{U}$(0.1, 1000.0) & $log\mathcal{U}$(0.1, 1000.0) \\
    $D_{H50/CTIOi}$ & Dilution factor for SDSS $i^{\prime}$ & - & $\mathcal{U}$(0.9, 1.0) & 1.0 \\
    $M_{H50/CTIOi}$ & Flux offset for SDSS $i^{\prime}$ &  - & $\mathcal{N}$(0.0, 0.1) & $\mathcal{N}$(0.0, 0.1) \\
    $J_{H50/CTIOi}$ [ppm] & Photometric jitter for SDSS $i^{\prime}$ & - & $log\mathcal{U}$(0.1, 1000.0) &  $log\mathcal{U}$(0.1, 1000.0) \\
    $\mu_{CHRION/FEROS}$ [m~s$^{-1}$] & RV zeropoint for CHIRON or FEROS & $\mathcal{N}$(49600, 200) & - & $\mathcal{N}$(33000, 200) \\
    $\mu_{HARPS}$ [m~s$^{-1}$] & RV zeropoint for HARPS & - & - & $\mathcal{N}$(33000, 200) \\
    $J_{CHIRON/FEROS}$ [m~s$^{-1}$] & RV jitter for CHIRON or FEROS & $log\mathcal{U}$(0.001, 100.0) & - & $log\mathcal{U}$(0.001, 100.0) \\
    $J_{HARPS}$ [m~s$^{-1}$] & RV jitter for HARPS & - & - & $log\mathcal{U}$(0.001, 100.0) \\
    \bottomrule
    \end{tabular}
    \caption{The priors used in the Juliet modelling of TASSIE II TOIs. Note:  $\mathcal{N}$, $\mathcal{U}$ and $log\mathcal{U}$ refer to normal, uniform and log uniform distributions, respectively.}
    \label{table:priors}
\end{table*}

\begin{figure*}
    \centering
    \includegraphics[width=0.95\linewidth, height=0.5\linewidth]{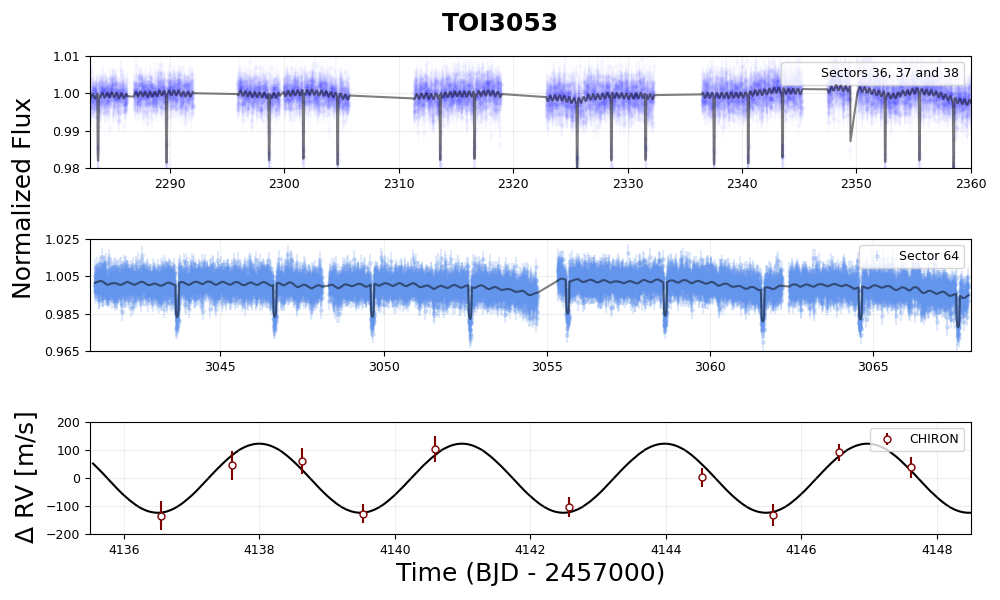}
    \caption{The \textit{TESS} multi-sector light curves and CHIRON radial velocities for TOI-3053, with best-fit transit + GP and RV models overlaid in black. In the transit light curves (top two panels), the different coloured points indicate the corresponding TESS sectors of the observations. The short-period, low amplitude signal is from the nearby contaminating eclipsing binary (discussed in Section 2).}
    \label{fig:TOI-3053}
\end{figure*}

\begin{figure*}
    \centering
    \includegraphics[width=0.95\linewidth, height=0.6\linewidth]{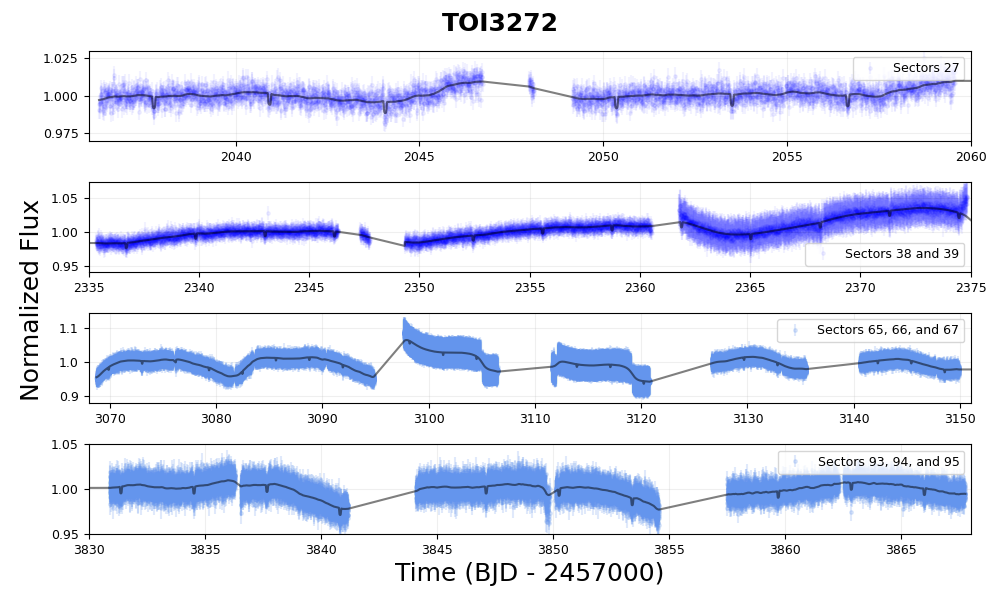}
    \caption{The multi-sector \textit{TESS} light curve for TOI-3272, with best-fit transit + GP model shown in black.}
    \label{fig:TOI-3272}
\end{figure*}

\begin{figure*}
    \centering
    \includegraphics[width=0.95\linewidth]{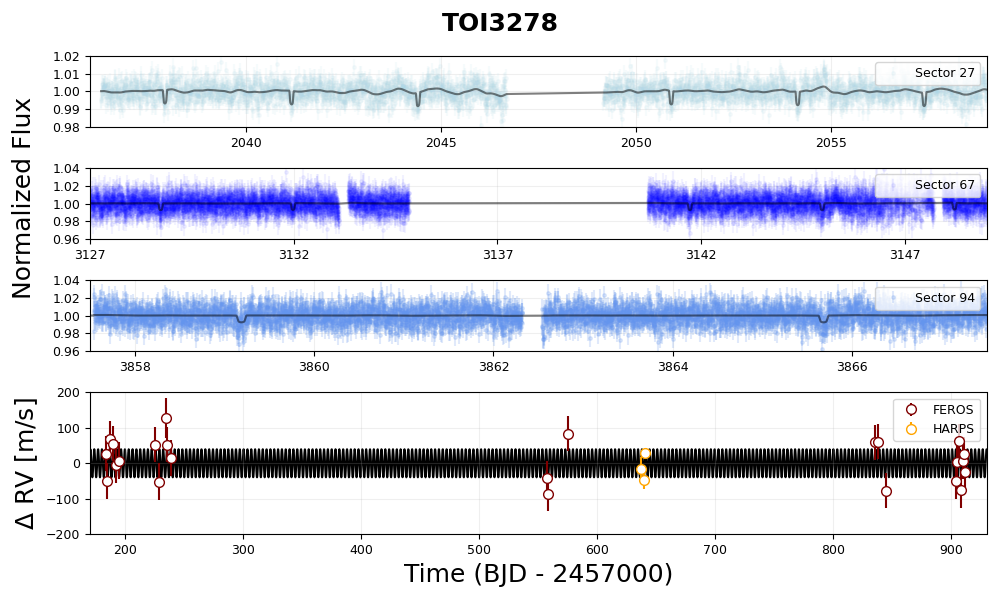}
    \caption{The \textit{TESS} multi-sector light curves and FEROS + HARPS radial velocities for TOI-3278, with best-fit transit + GP and RV models overlaid in black. In the transit light curves (top three panels), the different coloured points indicate the corresponding \textit{TESS} sectors for the observations. In the RV plot (bottom panel), the data are coloured by the spectrograph used for each measurement.}
    \label{fig:TOI-3278}
\end{figure*}

\begin{figure*}
     \begin{subfigure}[]
	   \centering
       \includegraphics[width=0.95\linewidth]{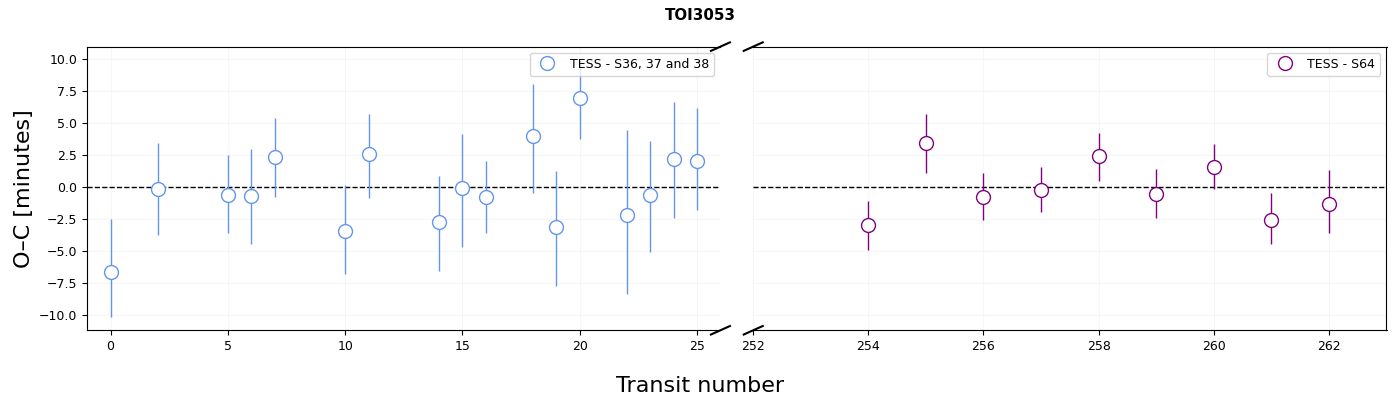}
    \end{subfigure}
    ~
    \begin{subfigure}[]
	   \centering
        \includegraphics[width=0.95\linewidth]{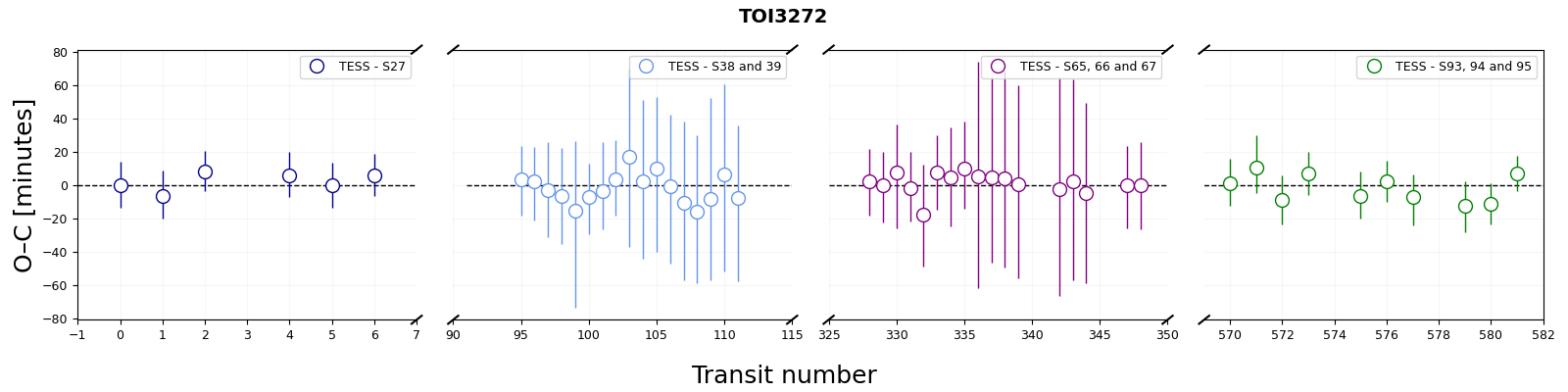}
    \end{subfigure}
    ~
    \begin{subfigure}[]
	   \centering
       \includegraphics[width=0.95\linewidth]{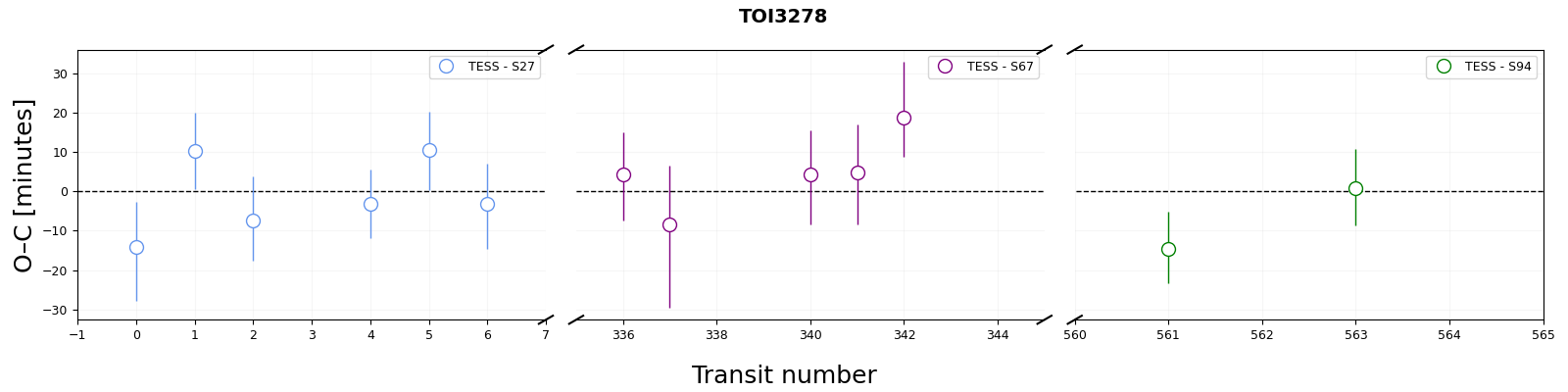}
    \end{subfigure}
    \caption{The O-C graphs for TOI-3053b \textbf{(a)}, TOI-3272.01 \textbf{(b)} and TOI-3278b / HATS-78b \textbf{(c)}. The error bars show the 1$\sigma$ (68 \% confidence band) errors on timing. The large uncertainties for TOI-3272 in Sectors 38 to 67 are a result of low signal-to-noise transits, due to the stellar activity and faintness of the host star.}
    \label{fig:TTVs}
\end{figure*}

\begin{figure*}
    \begin{subfigure}[]
	   \centering
       \includegraphics[width=0.45\linewidth]{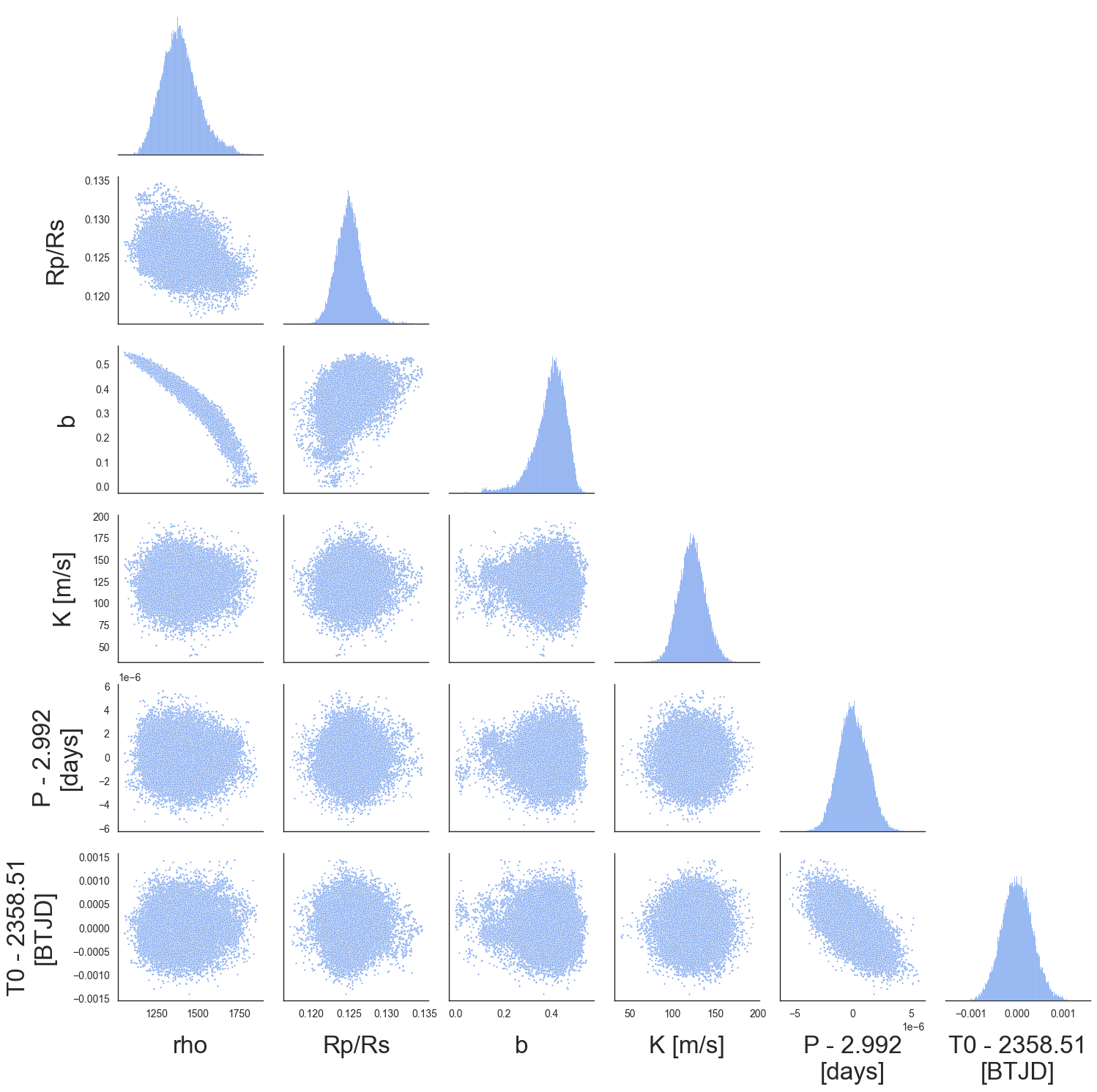}
    \end{subfigure}
    ~
    \begin{subfigure}[]
	   \centering
       \includegraphics[width=0.45\linewidth]{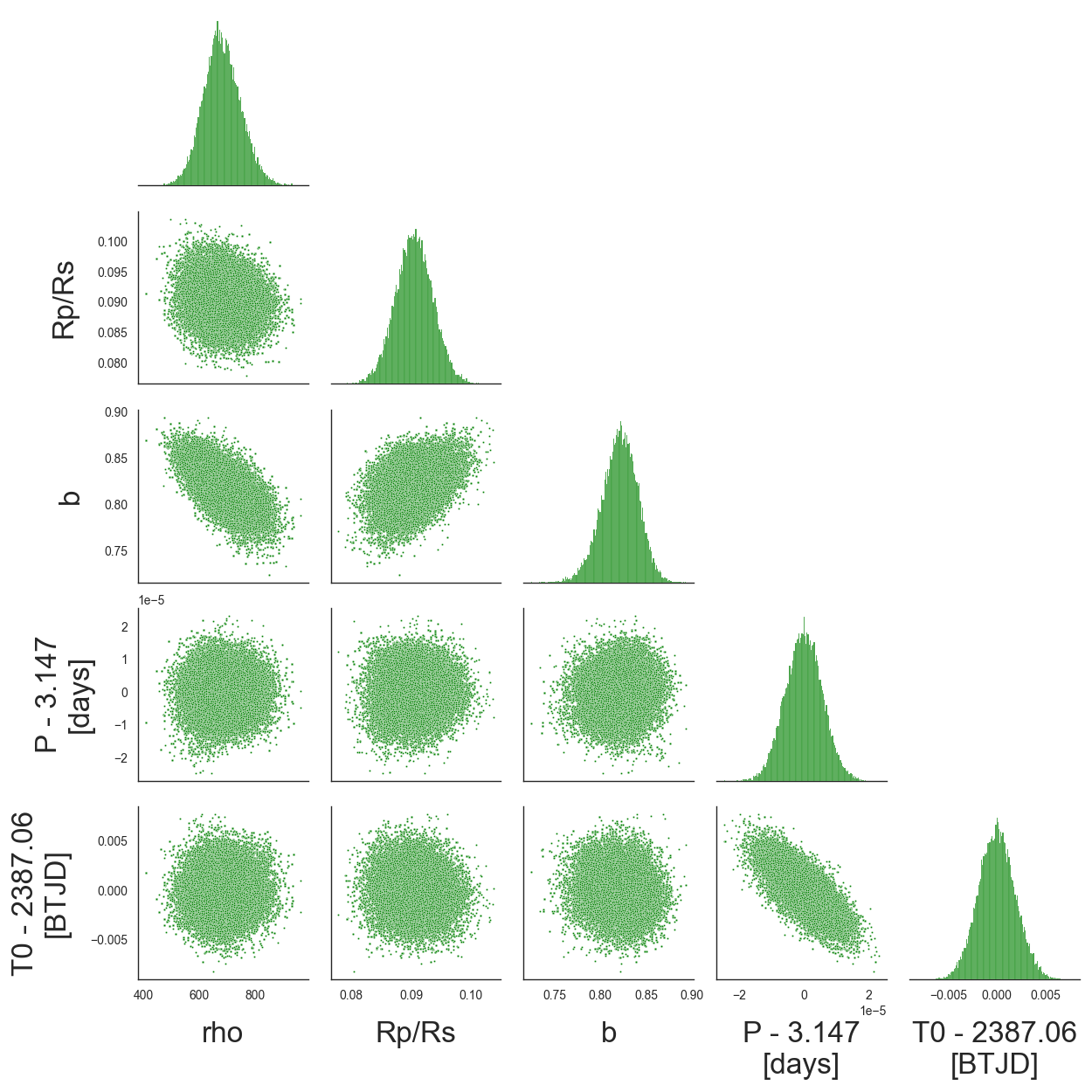}
    \end{subfigure}
    ~
    \begin{subfigure}[]
	   \centering
       \includegraphics[width=0.495\linewidth]{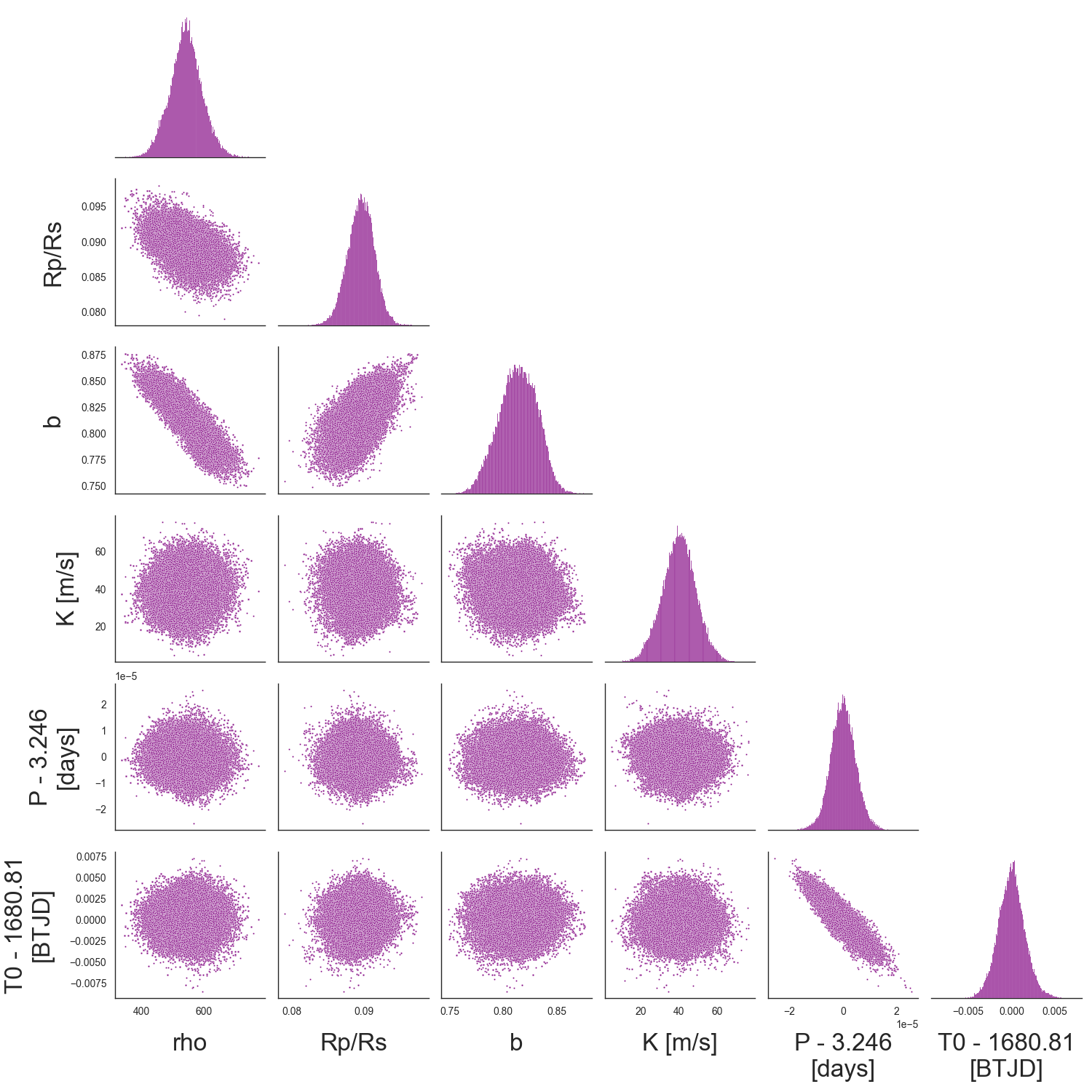}
    \end{subfigure}
    ~
    \caption{Corner plots of the posterior distributions from transit modelling with Juliet for TOI-3053b \textbf{(a)}, TOI-3272.01 \textbf{(b)} and TOI-3278b / HATS-78b \textbf{(c)}.}
    \label{fig:posteriors}
\end{figure*}

\begin{table*}[h]
	\begin{center}
		\begin{tabular}{|p{2.1cm}|p{5.1cm}|p{2.5cm}|p{1.8cm}|p{1.8cm}|p{1.8cm}|} 
			\toprule
			\headrow \textbf{Parameters} & \textbf{Description} & \textbf{Priors} & \textbf{TOI-3053} & \textbf{TOI-3272} & \textbf{TOI-3278}\\
            \midrule
			\headrow \multicolumn{6}{|c|}{\textbf{GP Parameters}} \\
            \midrule
            $Prot_{TESS}$ [days] & Rotation period for QPK kernel & See text\newline(Sections 3.4 \& 4.2)  & $0.4166^{+0.0001}_{-0.0001}$ & $13.44^{+0.46}_{-0.43}$ & - \\
            $B_{TESS600}$ [ppm] & Amplitude of the QPK GP (TESS 600 s) & $log\mathcal{U}(10^{-6}, 10^{+6})$ & $0.0008^{+0.0053}_{-0.0007}$ & $0.00017^{+0.00013}_{-0.00005}$ & - \\ 
            $C_{TESS600}$ & Additive factor of the QPK GP (TESS 600 s) & $log\mathcal{U}(10^{-6}, 10^{+6})$ & $2612.3^{+15914.4}_{-2118.0}$ & $0.002^{+0.085}_{-0.002}$ & - \\
            $L_{TESS600}$   & Length-scale of exponential part of the QPK GP (TESS 600 s) & $log\mathcal{U}(10^{-6}, 10^{+6})$ & $6145.0^{+40038.7}_{-5200.0}$ & $33.56^{+25.66}_{-11.47}$ & - \\
            $B_{TESS200}$ [ppm]  & Amplitude of the QPK GP for TESS (200 s) & $log\mathcal{U}(10^{-6}, 10^{+6})$ & $0.0009^{+0.0101}_{-0.0008}$ & $0.0005^{+0.0003}_{-0.0002}$ & - \\
            $C_{TESS200}$   & Additive factor of the QPK GP (TESS 200 s)  & $log\mathcal{U}(10^{-6}, 10^{+6})$ & $2809.5^{+39171.2}_{-2489.2}$ & $0.0003^{+0.0113}_{-0.0003}$ & - \\
            $L_{TESS200}$   & Length-scale of exponential part of the QPK GP (TESS 200 s) & $log\mathcal{U}(10^{-6}, 10^{+6})$ & $599.2^{+6593.0}_{-525.6}$ & $36.49^{+20.97}_{10.98}$ & - \\
            $\sigma_{TESS600}$ [ppm]  & Amplitude of the Matern 3/2 GP (TESS 600 s) & $log\mathcal{U}(10^{-6}, 10^{+8})$ & - & - & $0.0011^{+0.0001}_{-0.0001}$ \\
            $\rho_{TESS600}$    & Time-scale of the Matern 3/2 GP (TESS 600 s) & $log\mathcal{U}(10^{-3}, 10^{+5})$ & - & - & $0.1597^{+0.0356}_{-0.0306}$ \\
            $\sigma_{TESS200}$ [ppm] & Amplitude of the Matern 3/2 GP (TESS 200 s) & $log\mathcal{U}(10^{-6}, 10^{+8})$ & - & - & $0.0007^{+0.0010}_{-0.0003}$ \\
            $\rho_{TESS200}$    & Time-scale of the Matern 3/2 GP (TESS 200 s) & $log\mathcal{U}(10^{-3}, 10^{+5})$ & - & - & $0.2762^{+4824.7559}_{-0.1570}$ \\
            $\sigma_{H50r}$ [ppm] & Amplitude of the Matern 3/2 GP (SDSS $r^{\prime}$) & $log\mathcal{U}(10^{-6}, 10^{+6})$ & $0.06^{+0.23}_{-0.06}$ & $0.022^{+0.071}_{-0.016}$ & $0.0127^{+0.0627}_{-0.0115}$ \\
            $\rho_{H50r}$ & Time-scale of the Matern 3/2 GP (SDSS $r^{\prime}$) & $log\mathcal{U}(10^{-3}, 10^{+3})$ & $8.8^{+118.7}_{-8.7}$ & $0.143^{+0.277}_{-0.094}$ & $10.8782^{+134.5884}_{-10.1369}$ \\
            $\sigma_{H50g}$ [ppm]  & Amplitude of the Matern 3/2 GP (SDSS $g^{\prime}$) & $log\mathcal{U}(10^{-6}, 10^{+6})$ & $0.005^{+0.001}_{-0.001}$ & - & - \\
            $\rho_{H50g}$    & Time-scale of the Matern 3/2 GP (SDSS $g^{\prime}$) & $log\mathcal{U}(10^{-3}, 10^{+3})$ & $0.007^{+0.005}_{-0.003}$ & - & - \\ 
            $\sigma_{H50V}$ [ppm]  & Amplitude of the Matern 3/2 GP (Bessell V) & $log\mathcal{U}(10^{-6}, 10^{+6})$ & $0.02^{+0.12}_{-0.02}$ & $0.008^{+0.022}_{-0.004}$ & $0.0255^{+0.1563}_{-0.0215}$ \\
            $\rho_{H50V}$ & Time-scale of the Matern 3/2 GP (Bessell V) & $log\mathcal{U}(10^{-3}, 10^{+3})$ & $19.9^{+207.9}_{-19.0}$ & $0.082^{+0.167}_{-0.047}$ & $1.1491^{+6.5910}_{-0.9753}$ \\
            $\sigma_{H50/CTIOi}$ [ppm] & Amplitude of the Matern 3/2 GP (SDSS $i^{\prime}$) & $log\mathcal{U}(10^{-3}, 10^{+3})$ & - & $0.010^{+0.029}_{-0.005}$ & $0.0065^{+0.0420}_{-0.0057}$\\
            $\rho_{H50/CTIOi}$ & Time-scale of the Matern 3/2 GP (SDSS $i^{\prime}$) & $log\mathcal{U}(10^{-6}, 10^{+6})$ & - & $0.073^{+0.149}_{-0.039} $ & $1.5231^{+8.7437}_{-1.3812}$\\
			\bottomrule 
		\end{tabular}
		\caption{The GP priors and parameter estimates from modelling with Juliet.}
		\label{table:GPresults}
	\end{center}
\end{table*}



\end{document}